\documentclass[notitlepage,aps,prc,groupedaddress]{revtex4-1}
\usepackage{graphicx,amssymb,mathtools,float,subfig,hyperref,cleveref,braket,siunitx,enumerate}
\crefname{equation}{Eq.}{Eqs.}
\crefname{figure}{Fig.}{Figs.}
\crefname{table}{Table}{Tables}
\newcommand{\ccite}[1]{Ref.~\cite{#1}}

\newcommand{\bvec}[1]{\boldsymbol{#1}}

\newcommand{\dd}{\,\mathrm{d}}
\newcommand{\pp}{{\prime \prime}}
\newcommand{\up}[1]{{^{#1}}\!}
\newcommand{\units}[1]{\,\mathrm{#1}}
\begin{document}
\title{In-medium nucleon-nucleon cross-sections with non-spherical Pauli blocking}
\author{L. White}
\email[Email:]{whitel20ster@gmail.com}
\author{F. Sammarruca}

\email[Email:]{fsammarr@uidaho.edu}
\affiliation{Physics Department, University of Idaho, Moscow, ID 83844-0903, U.S.A}
\date{\today}

\begin{abstract}
We present a formalism to solve the Bethe-Goldstone scattering equation without the use of partial wave expansion  which is alternative to the one we developed in a previous work. The present approach is more suitable for the calculation of in-medium nucleon-nucleon cross sections, which are the focal point of this paper. The impact of removing the spherical approximation on the angle and energy dependence of, particularly,  in-medium proton-proton and proton-neutron differential cross sections is discussed along with its potential implication. 
\end{abstract}
\maketitle
\section{Introduction}

The Bethe-Goldstone equation~\cite{BLM,Bethe56,Goldstone,Bethe71} describes nucleon-nucleon (NN) scattering in
a dense hadronic medium. Conventional medium effects present in the equation include corrections of the single-particle energies
to account for the presence of the medium and the Pauli blocking mechanism, which prevents scattering into occupied
states. Within the Dirac-Brueckner-Hartree-Fock (DBHF) approach, which is our traditional framework, an additional ``non conventional" medium
effect comes in through the use of the (density-dependent) nucleon effective mass in the nucleon Dirac spinors.

In a previous work \cite{White2013}, we presented 
a method for the solution of the                      
Bethe-Goldstone equation without the use of partial wave decomposition in order to remove the spherical, or angle average, approximation on the Pauli blocking operator.  With that as our baseline, 
in this paper we consider the idealized scenario of two nucleons undergoing scattering in infinite, symmetric or
asymmetric, nuclear matter, and calculate in-medium cross sections for such a process, using as input a G-matrix
obtained from the solution of the Bethe-Goldstone equation in three-dimensional space. A crucial step for 
accomplishing such solution in a manageable way is the removal of the azimuthal degree of freedom from the 
starting three-dimensional equation. First, we will describe why we need to take a different approach than the 
one we adopted in Ref.~\cite{White2013} if we wish to 
place the incident momentum along the direction of the chosen quantization axis, the $z$-axis, as typically done
in standard calculations of NN observables. As a consequence of that, we will propose an alternative method to partially decouple the 
system of (helicity basis) equations. These and other technical issues (if significantly different than those previously reported), will be confronted in Section~\ref{Formalism}.

After incorporating (non-spherical) Pauli blocking and other appropriate medium effects, we proceed to
calculate in-medium differential cross sections.
Clearly in-medium scattering (that is, the scattering of two nucleons embedded in nuclear matter), is not a directly observable process.
A connection with physical scattering can be made considering, for instance, a nucleon bound in a
nucleus (or, more ideally, in nuclear matter, as in our case) through the nuclear mean
field. If such nucleon is struck [for instance, as in a ($e,e'$) reaction], it may
subsequently scatter from another nucleon. This process would require the knowledge of the
in-medium NN cross-section, or {\it effective} cross-section.

Another scenario which involves in-medium two-body cross-sections
is the dynamics of heavy-ion collisions. These are typically handled
with so-called transport equations, such as the Boltzmann-Uehling-Uhlenbeck equation \cite{BUU1,BUU2},
which describe the evolution of a system of strongly interacting hadrons
drifting in the presence of a mean field while undergoing two-body collisions.

In-medium cross-sections are driven by the scattering amplitudes as well as kinematical
factors. In a microscopic approach, they are constructed from the
(medium-modified) NN amplitudes without phenomenology.
They depend on several variables, such as the relative momentum of the two-nucleon
pair, the total momentum of the pair in the nuclear matter rest frame (needed for the Pauli operator), and,
potentially, two different densities or Fermi momenta.
To facilitate applications in reactions, these multiple dependences have been handled in different ways and
with different levels of approximations.
In the simplest approach, the assumption is made that the transition matrix in the medium is approximately
the same as in vacuum, and that medium effects on the cross-section come in only through the use of
nucleon effective masses in the phase space factors \cite{Pand,Pers,Li05}.
Concerning microscopic approaches, some can be found, for instance, in Refs.~\cite{LM,LM2,FFE}, but consideration
of medium asymmetries are not included in those predictions. Furthermore, in Refs.~\cite{LM,LM2}, the full
(complex) nature of the scattering amplitude is not taken into account, and in-medium differential cross sections
are calculated from a (real) $K$-matrix. Finally, all previous microscopic predictions make use of the 
angle-averaged Pauli operator.

Effective cross-sections can also provide information on the nucleon mean-free path in nuclear matter, a 
quantity of fundamental importance to understand the dynamics of a nucleon in the medium. 
In summary, they are an important input for several processes.
It is one of the goals of this paper to investigate to which extent our microscopic in-medium observables
are sensitive to the exact treatment of Pauli blocking, and, more generally, to examine how their angular
structure  and energy dependence are impacted by the medium. 

Our results are presented and discussed in Sec.~\ref{RD}, while Sec.~\ref{Summary} contains a short summary and
our conclusions. 

\section{Formal Aspects}\label{Formalism}
\subsection{Free space: the Thompson equation in a helicity basis}
As we did previously, we start with  the (free-space) Thompson equation in a helicity. More explicitly, the Thompson equation which we adopt is given as \cite{White2013}
\begin{equation}
\braket{\lambda_1^\prime \lambda_2^\prime|\hat{T}^I(\bvec{q}^\prime,\bvec{q})|\lambda_1 \lambda_2} = \braket{\lambda_1^\prime \lambda_2^\prime|\hat{V}^I(\bvec{q}^\prime,\bvec{q})|\lambda_1 \lambda_2} + \sum_{\lambda_1^\pp, \lambda_2^\pp = \pm } \int_{\mathbb{R}^3} \frac{\braket{\lambda_1^\prime \lambda_2^\prime|\hat{V}^I(\bvec{q}^\prime,\bvec{q}^\pp)|\lambda_1^\pp \lambda_2^\pp} \braket{\lambda_1^\pp \lambda_2^\pp|\hat{T}^I(\bvec{q}^\pp,\bvec{q})|\lambda_1 \lambda_2}}{2( E_q - E_{q^\pp} + i \epsilon)} \dd^3 q^\pp \; .
\label{Eq:Thompson_heli}
\end{equation}

Above $\hat{T}$, $\hat{V}$, $I$, $E_p = \sqrt{p^2 + m^2}$ and $m$ are the $T$-matrix, NN potential (see Appx. A of Ref.~\cite{White2013}), total isospin, relativistic energy, and nucleon mass (average of the proton and neutron mass) respectively. Furthermore, $\bvec{q}$, $\bvec{q}^\prime$, and $\bvec{q^\pp}$ are the initial, final, and intermediate momenta in the 
two-nucleon center-of-mass frame.

The first issue to be addressed is the removal of the azimuthal degree of freedom, which can be done in a number of ways.  The average over the azimuthal angle performed as in Eqs.~(7a-b) of Ref.~\ccite{White2013} in the spirit of Ref.~\cite{RK} causes some of the matrix elements to vanish
if the initial momentum is along the $z$-axis (i.e. $\theta=0$),  therefore here we will adopt a different prescription. Placing the initial momentum along the $z$-axis, we observe the resulting symmetry of the NN potential:
\begin{equation}
\braket{\lambda_1^\prime \lambda_2^\prime|\hat{V}^I(\bvec{q}^\prime,q,0,\phi)|\lambda_1 \lambda_2} = e^{i \Lambda(\phi^\prime - \phi)} \braket{\lambda_1^\prime \lambda_2^\prime|\hat{V}^I(\tilde{q}^\prime,0,q,0,0)|\lambda_1 \lambda_2} \equiv e^{i \Lambda(\phi^\prime - \phi)} \braket{\lambda_1^\prime \lambda_2^\prime|v^I(\tilde{q}^\prime,q)|\lambda_1 \lambda_2} \; ,
\end{equation}
with $\Lambda \equiv \lambda_1 - \lambda_2$ and $\tilde{q}^\prime \equiv (q^\prime,\theta^\prime)$. This symmetry, which carries over to the $T$-matrix, can easily be shown by writing the $T$-matrix (and NN potential) in a partial wave helicity basis expansion [see \cref{Eq:transformation}].

After setting $\theta=0$ in \cref{Eq:Thompson_heli} and implementing the previous observation, in the spirit of Ref.~\cite{FEG} we apply on both side of the equation the operator $\frac{1}{2 \pi} \int_0^{2 \pi} e^{-i \Lambda (\phi^\prime - \phi)} \dd \phi^\prime$ and arrive at the $\phi$-integrated Thompson equation
\begin{align}
\braket{ \lambda_1^\prime \lambda_2^\prime  |t^I(\tilde{q}^\prime,q)| \lambda_1 \lambda_2 } & = \braket{ \lambda_1^\prime \lambda_2^\prime  |v^I(\tilde{q}^\prime,q)| \lambda_1 \lambda_2 } \nonumber \\
&+ \sum_{\lambda_1^\pp, \lambda_2^\pp = \pm } \pi \int_0^\infty \int_0^\pi \frac{\braket{ \lambda_1^\prime \lambda_2^\prime  | v^{\Lambda I}(\tilde{q}^\prime,\tilde{q}^\pp) | \lambda_1^\pp \lambda_2^\pp } \braket{ \lambda_1^\pp \lambda_2^\pp  |t^I(\tilde{q}^\pp,q)| \lambda_1 \lambda_2 } } {E_q - E_{q^\pp} + i \epsilon} {q^\pp}^2 \sin \theta^\pp \dd \theta^\pp \dd q^\prime \; ,
\label{Eq:phi_integrated_thompson}
\end{align}
with the \textit{real} $\phi$-integrated potential equal to
\begin{equation}
\braket{ \lambda_1^\prime \lambda_2^\prime  | v^{\Lambda I}(\tilde{q}^\prime,\tilde{q}^\pp) | \lambda_1^\pp \lambda_2^\pp } = \frac{1}{2 \pi} \int_0^{2 \pi} e^{i \Lambda (\phi^\pp - \phi^\prime)} \braket{ \lambda_1^\prime \lambda_2^\prime  | \hat{V}^I(\bvec{q}^\prime,\bvec{q}^\pp) | \lambda_1^\pp \lambda_2^\pp }|_{\phi^\prime = 0} \dd \phi^\pp \; .
\label{Eq:phipotential}
\end{equation}
Note that the $\phi$-integrated potential depends on double, single, and unprimed helicities, rendering the present set of equations different than Eq.~(8) of Ref.~\cite{White2013}. This calls for a different strategy to partially decouple the 
system.

\subsection{Partially decoupling the system of integral equations}
The $\phi$-integrated Thompson equations are a set of sixteen coupled Fredholm integral equations of the second kind for each isospin. Due to parity and isospin conservation, only six amplitudes are independent. For the six independent amplitudes we choose~\cite{White2013}
\begin{alignat}{3}
t^I_1 & \equiv \braket{++|t^I|++} \; , & \quad t^I_2 & \equiv \braket{++|t^I|--} \; , & \quad t^I_3 & \equiv \braket{+-|t^I|+-} \; , \nonumber \\
t^I_4 & \equiv \braket{+-|t^I|-+} \; , & t^I_5 & \equiv \braket{++|t^I|+-} \; , & t^I_6 & \equiv \braket{+-|t^I|++} \; .
\label{Eq:sixamplitudes}
\end{alignat}
Additionally, we found that the following linear combinations
\begin{equation}
\up{0}t^I \equiv t^I_1 - t^I_2 \; ,
\quad
\up{12}t^I \equiv t^I_1 + t^I_2 \; ,
\label{Eq:linearcombinations}
\end{equation}
will partially decouple the system for the  $t^I_1$ and $t^I_2$ amplitudes.

The symmetries on the $\phi$-integrated potential can be found analytically (for instance using a partial wave helicity basis expansion [see \cref{Eq:transformation}], or numerically. Either way,  we end up with the following conclusions: 1) When $\Lambda = 0$ the symmetries of the $\phi$-integrated potential are the same as the corresponding ones for the $t$-matrix, see Eq.~(9) of Ref.~\cite{White2013}. 2) When $\Lambda = \pm 1$ the following holds
\begin{align}
v_1^{1I} &\equiv \braket{++|v^{1 I}|++} = \braket{++|v^{-1 I}|++} = \braket{--|v^{1 I}|--}  = \braket{--|v^{-1 I}|--} \; , \nonumber \\
v_2^{1I} &\equiv \braket{++|v^{1 I}|--} = \braket{++|v^{-1 I}|--} = \braket{--|v^{1 I}|++} = \braket{--|v^{-1 I}|++}  \; , \nonumber \\
v_3^{1I} &\equiv \braket{+-|v^{1 I}|+-} = \braket{-+|v^{-1 I}|-+} \; , \nonumber \\
v_3^{-1I} &\equiv \braket{+-|v^{-1 I}|+-} = \braket{-+|v^{1 I}|-+} \; , \nonumber \\
v_4^{1I} &\equiv \braket{+-|v^{1 I}|-+} = \braket{-+|v^{-1 I}|+-} \; , \nonumber \\
v_4^{-1I} &\equiv \braket{+-|v^{-1 I}|-+} = \braket{-+|v^{1 I}|+-} \; , \nonumber \\
v_5^{1I} &\equiv \braket{++|v^{1 I}|+-} = -\braket{++|v^{-1 I}|-+} = \braket{--|v^{1 I}|+-} = -\braket{--|v^{-1 I}|-+} \; , \nonumber \\
v_5^{-1I} &\equiv \braket{++|v^{-1 I}|+-} = -\braket{++|v^{1 I}|-+} =  -\braket{--|v^{1 I}|-+} = \braket{--|v^{-1 I}|+-} \; , \nonumber \\
v_6^{1I} &\equiv \braket{+-|v^{1 I}|++} = -\braket{-+|v^{-1 I}|++} = \braket{+-|v^{1 I}|--} = -\braket{-+|v^{-1 I}|--} \; , \nonumber \\
v_6^{-1I} &\equiv \braket{+-|v^{-1 I}|++} = -\braket{-+|v^{1 I}|++} = -\braket{-+|v^{1 I}|--} = \braket{+-|v^{-1 I}|--}  \; .
\label{Eq:lambda_potential}
\end{align}

If we utilize \cref{Eq:phi_integrated_thompson,Eq:sixamplitudes,Eq:linearcombinations,Eq:lambda_potential} along with the symmetries of the $t$-matrix (see Eq.~9 in Ref.~\cite{White2013}) we obtain the following six partially coupled integral equations.

The spin singlet amplitude $^0t^I$ is uncoupled
\begin{equation}
{^0}t^I(\tilde{q}^\prime,q)  = {^0}v^I(\tilde{q}^\prime,q) + \pi \int_0^\infty \int_0^\pi \frac{ {^0}v^{0I}(\tilde{q}^\prime,\tilde{q}^\pp) {^0}t^I(\tilde{q}^\pp,q)}{E_q - E_q^\pp +i \epsilon} {q^\pp}^2 \sin \theta^\pp \dd \theta^\pp \dd q^\pp \; .
\end{equation}

The spin triplet amplitudes $^{12}t^I$ and $^{66}t^I$ form a bi-coupled system
\begin{align}
{^{12}}t^I(\tilde{q}^\prime,q)  &= {^{12}}v^I(\tilde{q}^\prime,q) +  \pi \int_0^\infty \int_0^\pi \frac{ {^{12}}v^{0I}(\tilde{q}^\prime,\tilde{q}^\pp) {^{12}}t^I(\tilde{q}^\pp,q)+ 4 v_5^{0I}(\tilde{q}^\prime,\tilde{q}^\pp) t_6^I(\tilde{q}^\pp,q) }{E_q - E_q^\pp +i \epsilon} {q^\pp}^2 \sin \theta^\pp \dd \theta^\pp \dd q^\pp \; , \\
t_6^I(\tilde{q}^\prime,q) &= v_6^I(\tilde{q}^\prime,q) +  \pi \int_0^\infty \int_0^\pi \frac{ v_6^{0I}(\tilde{q}^\prime,\tilde{q}^\pp) {^{12}}t^I(\tilde{q}^\pp,q) + [v_3^{0I} (\tilde{q}^\prime,\tilde{q}^\pp)-v_4^{0I}(\tilde{q}^\prime,\tilde{q}^\pp)] t_6^I(\tilde{q}^\pp,q)}{E_q - E_q^\pp +i \epsilon} {q^\pp}^2 \sin \theta^\pp \dd \theta^\pp \dd q^\pp \; .
\end{align}

Finally, the $t^I_3$, $t^I_4$, and $t^I_5$ amplitudes form a tri-coupled system
\begin{align}
t_3^I(\tilde{q}^\prime,q)  &= v_3^I(\tilde{q}^\prime,q) +  \pi \int_0^\infty \int_0^\pi \frac{ v_3^{1I}(\tilde{q}^\prime,\tilde{q}^\pp) t_3^I(\tilde{q}^\pp,q) + v_4^{1I}(\tilde{q}^\prime,\tilde{q}^\pp)t_4^I(\tilde{q}^\pp,q) + 2 v_6^{1I}(\tilde{q}^\prime,\tilde{q}^\pp)  t_5(\tilde{q}^\pp,q)}{E_q - E_q^\pp +i \epsilon} {q^\pp}^2 \sin \theta^\pp \dd \theta^\pp \dd q^\pp \; , \\
t_4^I(\tilde{q}^\prime,q)  &= v_4^I(\tilde{q}^\prime,q) +  \pi \int_0^\infty \int_0^\pi \frac{ v_4^{-1I}(\tilde{q}^\prime,\tilde{q}^\pp) t_3^I(\tilde{q}^\pp,q) + v_3^{-1I}(\tilde{q}^\prime,\tilde{q}^\pp)t_4^I(\tilde{q}^\pp,q) - 2 v_6^{-1I}(\tilde{q}^\prime,\tilde{q}^\pp) t_5(\tilde{q}^\pp,q)}{E_q - E_q^\pp +i \epsilon} {q^\pp}^2 \sin \theta^\pp \dd \theta^\pp \dd q^\pp \; , \\
t_5^I(\tilde{q}^\prime,q)  &= v_5^I(\tilde{q}^\prime,q) +  \pi \int_0^\infty \int_0^\pi \frac{ v_5^{1I}(\tilde{q}^\prime,\tilde{q}^\pp) t_3^I(\tilde{q}^\pp,q) - v_5^{-1I}(\tilde{q}^\prime,\tilde{q}^\pp)t_4^I(\tilde{q}^\pp,q) + {^{12}}v^{1I}(\tilde{q}^\prime,\tilde{q}^\pp) t_5(\tilde{q}^\pp,q)}{E_q - E_q^\pp +i \epsilon}{q^\pp}^2 \sin \theta^\pp \dd \theta^\pp \dd q^\pp  \; .
\end{align}
Concerning the numerical solution,  the strategies shown in Appx. B of \ccite{White2013} can be easily adapted to this particular set of equations.

\subsection{Connection with partial wave decomposition and construction of NN observables} \label{pwd}
Although the point of this paper is to avoid the method of partial-wave expansion, we utilize the partial wave solution for comparison purposes. The expansion of $\hat{T}^I(\bvec{q}^\prime,\bvec{q})$ in a partial wave helicity basis \cite{JW59,BrJack} is given by
\begin{equation}
\braket{\lambda_1^\prime \lambda_2^\prime|\hat{T}^I(\bvec{q}^\prime,\bvec{q})|\lambda_1 \lambda_2} = \sum_{JM} \frac{2J+1}{4 \pi} D^J_{M \Lambda^\prime}(\phi^\prime,\theta^\prime,-\phi^\prime)^\ast \braket{\lambda_1^\prime \lambda_2^\prime|\hat{T}^{IJ}(q^\prime,q)|\lambda_1 \lambda_2} D^J_{M \Lambda}(\phi,\theta,-\phi) \; ,
\label{Eq:transformation}
\end{equation}
where the Wigner $D$-matrix $D^J_{M \Lambda}(\alpha,\beta,\gamma) = e^{-iM \alpha} d^J_{M \Lambda}(\beta) e^{-i \Lambda \gamma}$ includes the reduced rotation matrix $d^J_{M \Lambda}(\beta)$ with $\Lambda \equiv \lambda_1-\lambda_2$ and an analogous definition for the primed coordinate. The partial wave amplitudes, denoted by $\hat{T}^{IJ}(q^\prime,q)$ (with a similar decomposition done for the NN potential), are the solutions of the partial wave decomposed Eq.~(\ref{Eq:Thompson_heli}).

To obtain a transformation from partial waves into the (angle-dependent) $t$-matrix we evaluate \cref{Eq:transformation} at $ \phi^\prime = \theta = \phi = 0$
\begin{equation}
\braket{ \lambda_1^\prime \lambda_2^\prime  |t^I(\tilde{q}^\prime,q)| \lambda_1 \lambda_2 } =  \sum_J \frac{2J+1}{4 \pi}  d^J_{\Lambda \Lambda^\prime} (\theta^\prime) \braket{ \lambda_1^\prime \lambda_2^\prime  |\hat{T}^{IJ}(q^\prime,q)| \lambda_1 \lambda_2 } \; .
\label{Eq:tmatunphy}
\end{equation}

As it stands, our angle-dependent solutions contain unphysical states. On the other hand, the well-known antisymmetry requirement for the NN system
imply that only even or odd values of $J$ are allowed in a particular state of definite spin and isospin.
Thus, starting with \cref{Eq:tmatunphy} and making use of the identities
 $(-1)^J d^J_{00}(\theta^\prime) = d^J_{00}(\pi-\theta^\prime)$, $(-1)^{J+1} d^J_{01}(\theta^\prime) = d^J_{01}(\pi-\theta^\prime)$, $(-1)^J d^J_{11} (\theta^\prime) = -d^J_{-11}(\pi-\theta^\prime)$, and $(-1)^J d^J_{-11} (\theta^\prime) = -d^J_{11}(\pi-\theta^\prime)$ we can, in each case, identify the appropriate combination of the direct and the exchange terms which must enter the antisymmetrized amplitudes.
For those, we obtain:
\begin{equation}
{^0}t_a^{\overset{1}{0}}(\tilde{q}^\prime,q) \equiv \frac{1}{2}[{^0}t^{\overset{1}{0}}(\tilde{q}^\prime,q) \pm {^0}t^{\overset{1}{0}}(-\tilde{q}^\prime,q)] = \sum_{J= \substack{\mathrm{even} \\ \mathrm{odd}}} \frac{2J+1}{4 \pi}  d^J_{00} (\theta^\prime) {^0} T^{J{\overset{1}{0}}}(q^\prime,q) \; ,
\end{equation}

\begin{equation}
{^{12}}t_a^{\overset{1}{0}}(\tilde{q}^\prime,q) \equiv \frac{1}{2}[{^{12}}t^{\overset{1}{0}}(\tilde{q}^\prime,q) \pm {^{12}}t^{\overset{1}{0}}(-\tilde{q}^\prime,q)] = \sum_{J= \substack{\mathrm{even} \\ \mathrm{odd}}} \frac{2J+1}{4 \pi}  d^J_{00} (\theta^\prime) {^{12}} T^{J{\overset{1}{0}}}(q^\prime,q) \; ,
\end{equation}
with,
\begin{equation}
t_{a,1}^{\overset{1}{0}}(\tilde{q}^\prime,q) = \frac{1}{2}[ {^{12}}t_a^{\overset{1}{0}}(\tilde{q}^\prime,q) + {^{0}}t_a^{\overset{1}{0}}(\tilde{q}^\prime,q) ] \qquad \mathrm{and} \qquad t_{a,2}^{\overset{1}{0}}(\tilde{q}^\prime,q) = \frac{1}{2}[ {^{12}}t_a^{\overset{1}{0}}(\tilde{q}^\prime,q) - {^{0}}t_a^{\overset{1}{0}}(\tilde{q}^\prime,q) ] \; ,
\end{equation}

\begin{equation}
t_{a,6}^{\overset{1}{0}}(\tilde{q}^\prime,q) \equiv \frac{1}{2}[t_6^{\overset{1}{0}}(\tilde{q}^\prime,q) \mp t_6^{\overset{1}{0}}(-\tilde{q}^\prime,q)] = \frac{1}{2}\sum_{J= \substack{\mathrm{even} \\ \mathrm{odd}}} \frac{2J+1}{4 \pi}  d^J_{01} (\theta^\prime) {^{66}} T^{J{\overset{1}{0}}}(q^\prime,q) \; ,
\end{equation}

\begin{equation}
t_{a,3}^{\overset{1}{0}}(\tilde{q}^\prime,q) \equiv \frac{1}{2}[t_3^{\overset{1}{0}}(\tilde{q}^\prime,q) \mp t_4^{\overset{1}{0}}(-\tilde{q}^\prime,q)] = \frac{1}{2} \left[ \sum_{J= \substack{\mathrm{even} \\ \mathrm{odd}}} \frac{2J+1}{4 \pi}  d^J_{11} (\theta^\prime) {^{34}} T^{J{\overset{1}{0}}}(q^\prime,q) + \sum_{J= \substack{\mathrm{odd} \\ \mathrm{even}}} \frac{2J+1}{4 \pi}  d^J_{11} (\theta^\prime) {^{1}} T^{J{\overset{1}{0}}}(q^\prime,q) \right] \; ,
\end{equation}
\begin{equation}
t_{a,4}^{\overset{1}{0}}(\tilde{q}^\prime,q) \equiv \frac{1}{2}[t_4^{\overset{1}{0}}(\tilde{q}^\prime,q) \mp t_3^{\overset{1}{0}}(-\tilde{q}^\prime,q)] = \frac{1}{2} \left[ \sum_{J= \substack{\mathrm{even} \\ \mathrm{odd}}} \frac{2J+1}{4 \pi}  d^J_{-11} (\theta^\prime) {^{34}} T^{J{\overset{1}{0}}}(q^\prime,q) - \sum_{J= \substack{\mathrm{odd} \\ \mathrm{even}}} \frac{2J+1}{4 \pi}  d^J_{-11} (\theta^\prime) {^{1}} T^{J{\overset{1}{0}}}(q^\prime,q) \right] \; ,
\end{equation}

\begin{equation}
t_{a,5}^{\overset{1}{0}}(\tilde{q}^\prime,q) \equiv \frac{1}{2}[t_5^{\overset{1}{0}}(\tilde{q}^\prime,q) \mp t_5^{\overset{1}{0}}(-\tilde{q}^\prime,q)] = \frac{1}{2}\sum_{J= \substack{\mathrm{even} \\ \mathrm{odd}}} \frac{2J+1}{4 \pi}  d^J_{10} (\theta^\prime) {^{55}} T^{J{\overset{1}{0}}}(q^\prime,q) \; ,
\end{equation}
where definitions for the linear combinations of partial wave amplitudes $\up{n}T^{J{\overset{1}{0}}} ,\; n = 0,1,12,34,55,66$ are given in \ccite{Mac87}. Also, above one must read across the top (or bottom) to associate the correct sign with the appropriate $J$ values (even or odd) and isospin ($0$ or $1$). We also used the shorthand notation for the exchange amplitude $t^I(-\tilde{q}^\prime,\tilde{q})=t^I(q^\prime,\pi-\theta^\prime,q,\theta)$.

We are now in a position to calculate observables as functions of the scattering angle relative to the $z$-axis. More technical details can be found in Ref.~\cite{LW2014}.

\subsection{Including the Pauli operator}\label{BG}
For completeness, we recall that, 
in analogy with the free-space case, the Bethe-Goldstone equation can be written
as~\cite{White2013}
\begin{equation}
\hat{G}^I(\bvec{q}^\prime,\bvec{q},\bvec{P},k_F) = \hat{V}^I(\bvec{q}^\prime,\bvec{q}) + \int_{\mathbb{R}^3} \frac{\hat{V}^I(\bvec{q}^\prime,\bvec{q}^\pp) Q(\bvec{q}^\pp,\bvec{P},k_F) \hat{G}^I(\bvec{q}^\pp,\bvec{q},\bvec{P},k_F)}{2(E^\ast_q - E^\ast_{q^\pp} + i \epsilon)} \dd^3 q^\pp \; ,
\label{Eq:BG_full}
\end{equation}
where the asterix signifies in-medium energies and the Pauli operator $Q$ suppresses scattering into states below the Fermi momentum. More explicitly the Pauli operator for symmetric nuclear matter is defined as
\begin{equation}
Q(\bvec{q}^\pp,\bvec{P},k_F) \equiv \Theta( |\bvec{P} + \bvec{q}^\pp| - k_F ) \Theta( |\bvec{P} - \bvec{q}^\pp| - k_F ) \; ,
\label{Eq:Q}
\end{equation}
where $\Theta$ is the Heaviside step function, $\bvec{P}$ is one half the center of mass momentum,
$\bvec{P} \pm \bvec{q}$ are the momenta of the two nucleons in the nuclear matter rest frame,
and $k_F$ is the Fermi momentum, related to the nucleon density by $\rho=\frac{2k_F^3}{3 \pi^2}$. Clearly, the free-space equation is recovered by using free-space energies and setting
$Q$=1.
The corresponding 
$\phi$-integrated Bethe-Goldstone equation
can then be solved in perfect analogy with the free-space case described above. 

We also recall that the  so called spherical or angle-averaged Pauli operator $\bar{Q}$ (see Ref.~\cite{Haf} and references therein), is 
obtained from 
\begin{equation}
Q(\bvec{q}^\pp,\bvec{P},k_F) \approx \bar{Q}(q^\pp,P,k_F) = \frac{\int Q(\bvec{q}^\pp,\bvec{P},k_F) \dd \Omega^\pp }{ \int \dd \Omega^\pp } ,
\label{Eq:Q_avg}
\end{equation}
unless it's equal to zero or one. Clearly, no angle average is required in our three-dimensional approach.

The case of asymmetric nuclear matter, that is, when two different Fermi momenta, $k_{F1}$ and $k_{F2}$, are present (as would be the case in collisions of two different ions), can be handled by simply modifying the angular integration to implement the restrictions
\begin{align}
& |\bvec{P} + \bvec{q}^\pp|  > k_{F1} \quad \text{and} \quad  |\bvec{P} - \bvec{q}^\pp| > k_{F2} \implies \nonumber \\
& \quad Q(\bvec{q}^\pp,\bvec{P},k_{F1},k_{F2}) \equiv \Theta( |\bvec{P} + \bvec{q}^\pp| - k_{F1} ) \Theta( |\bvec{P} - \bvec{q}^\pp| - k_{F2} ) \; ,
\label{Eq:Q2}
\end{align}
which, again, is easily accomplished in our three-dimensional formalism.

\section{Results and Discussion}\label{RD}
The scattering amplitudes obtained
from the solution of the integral equations as described above are the input for calculating
NN scattering observables.
In this section, we will present and discuss selected {\it in-medium}
$np$ and $pp$ cross sections obtained with the exact Pauli operator and
compare with previous predictions which utilize the angle-averaged expression.
Once again, NN scattering in nuclear matter is not directly measurable, but a model  for such process can
be indirectly tested through applications in nuclear reactions.

In addition to the elastic differential cross-section, we will also consider polarized scattering,      
to explore whether the sensitivities we are investigating are more or less pronounced
in the spin dependence of the interaction. As a representative example, we have chosen the
 depolarization parameter, $D$, or $D_{nn}$, 
which refers to an experiment where the spin polarization normal to the scattering plane
is observed for the beam and the scattered particle.

\begin{figure}[H]
\begin{center}
\subfloat{
\includegraphics[width=7cm]{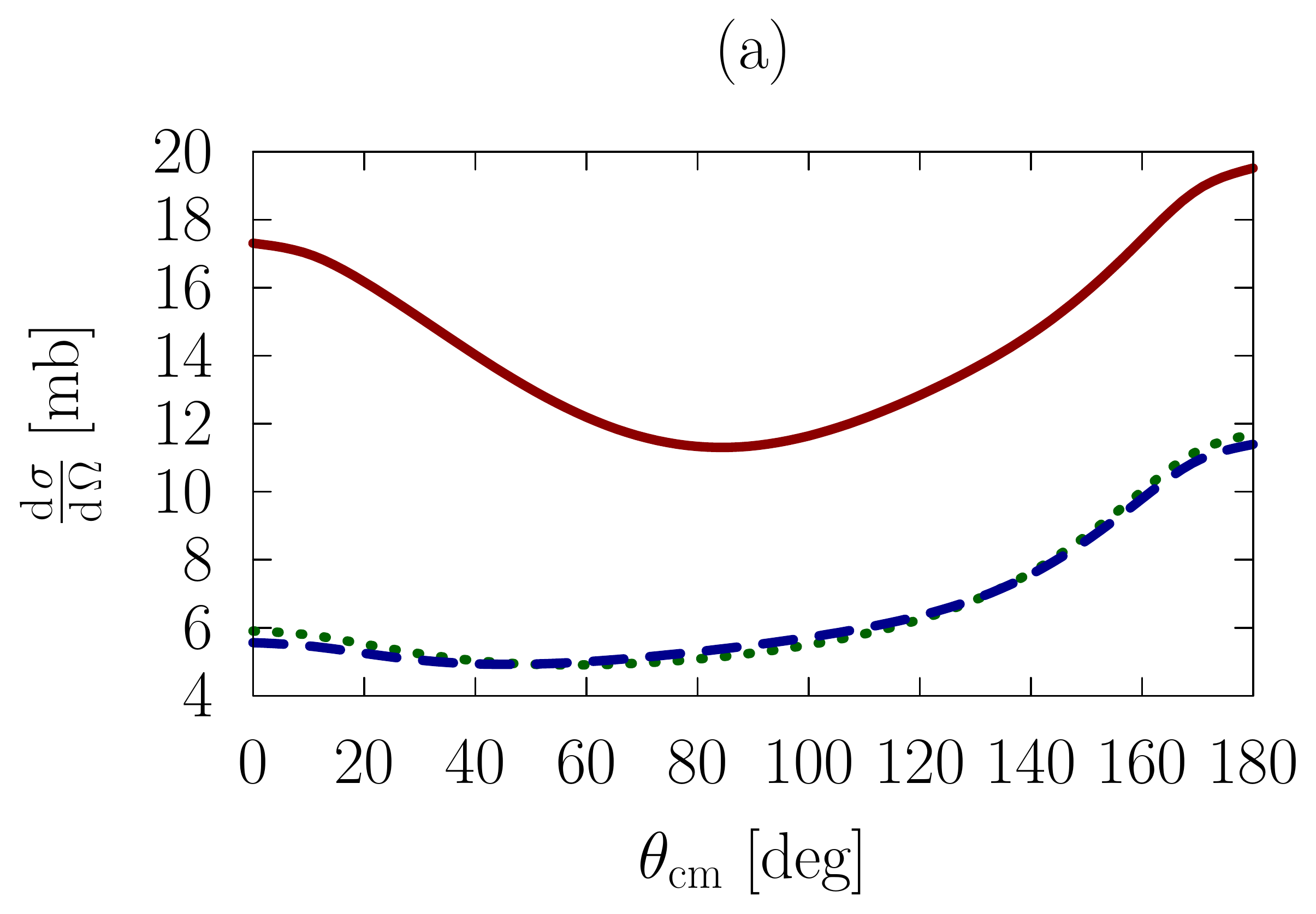}
}
\subfloat{
\includegraphics[width=7cm]{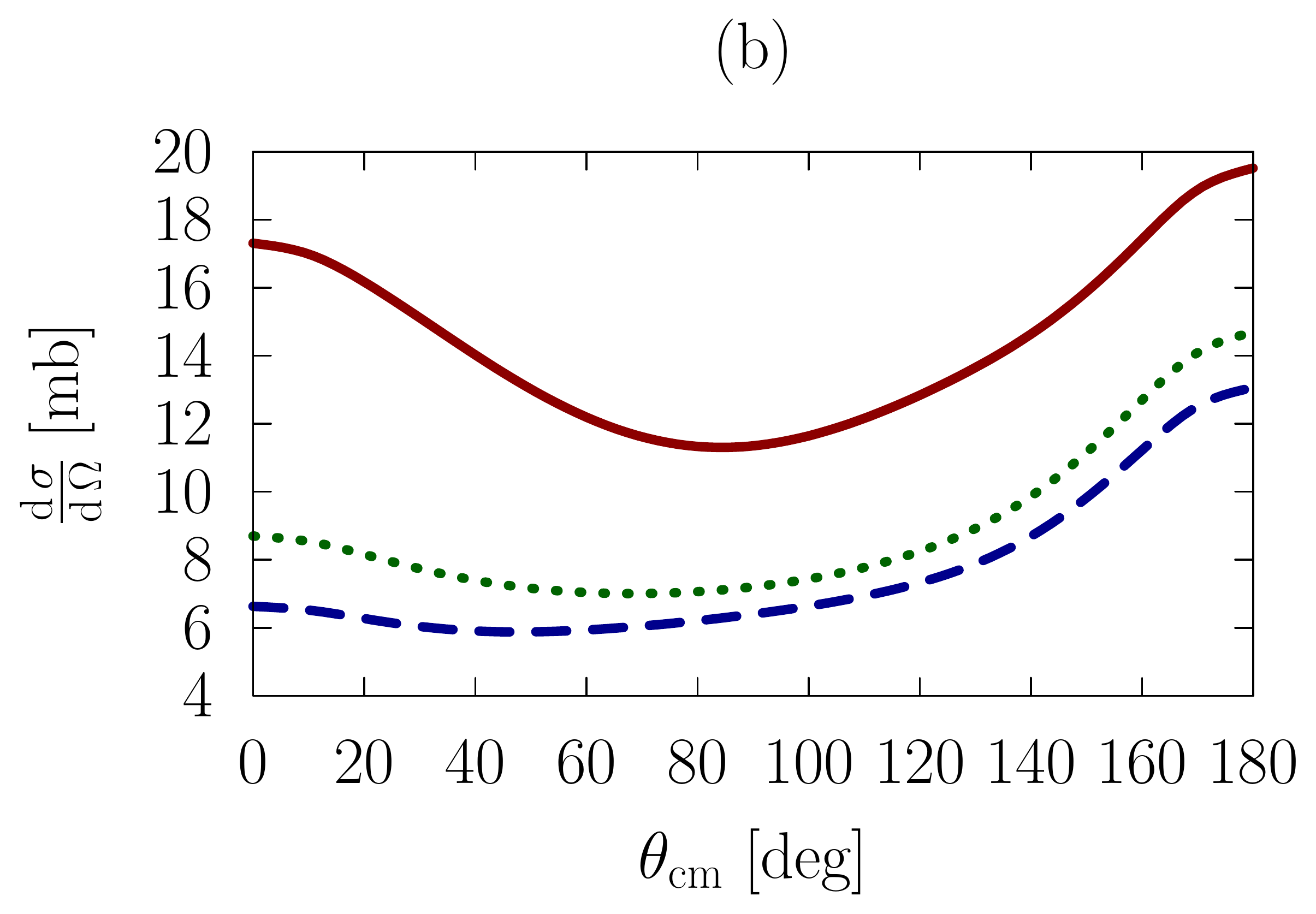}
}
\\
\subfloat{
\includegraphics[width=7cm]{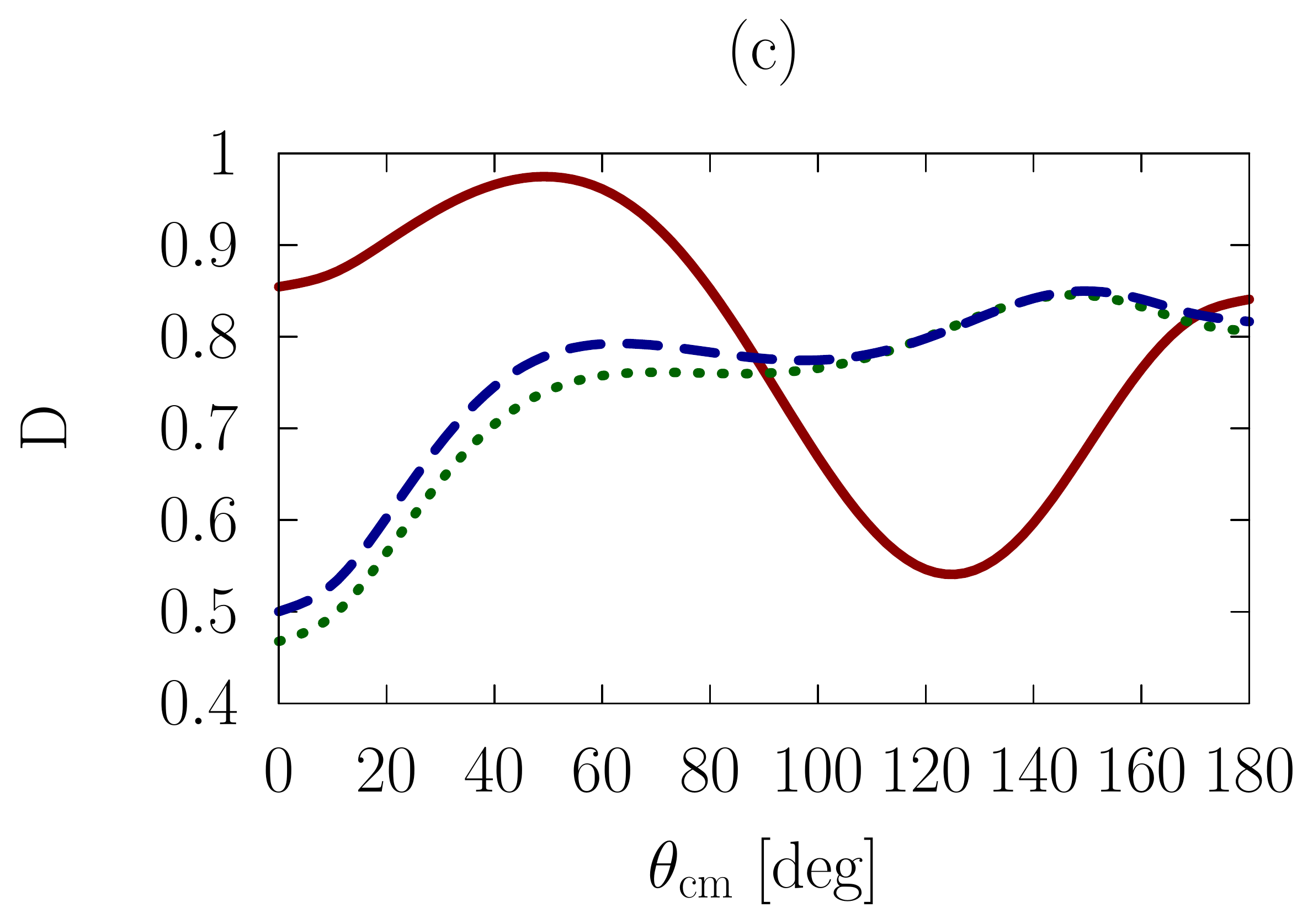}
}
\subfloat{
\includegraphics[width=7cm]{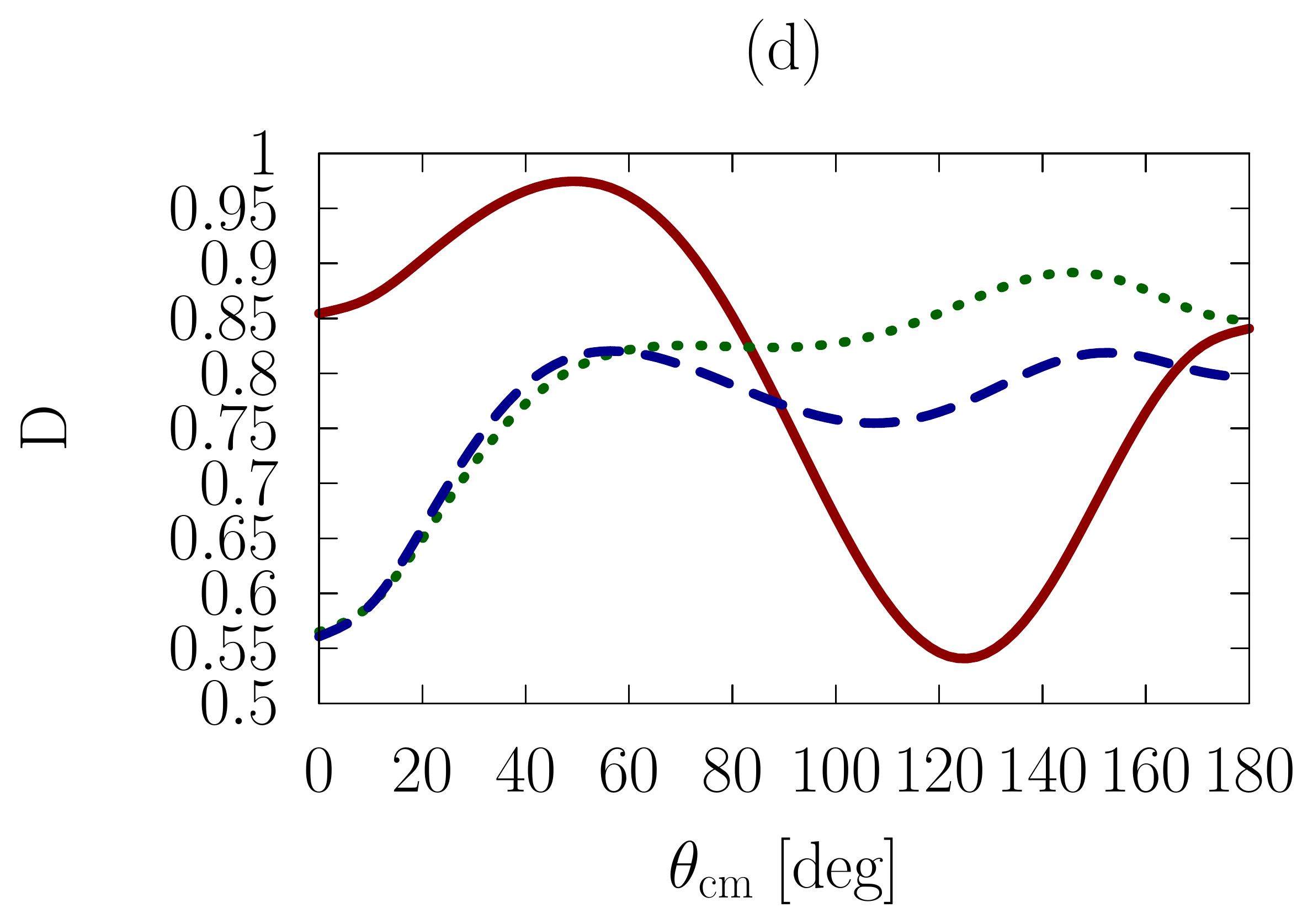}
}
\end{center}
\caption{(Color online) [Figs. (a) and (b)] $np$ elastic differential cross-section and [Figs. (c) and (d)] depolarization at a laboratory free-space
energy of 50 MeV {\it vs.} the c.m. scattering angle. The solid red curve shows the free-space prediction. The angle-averaged calculation is given by the dotted green curve whereas the dashed blue curve shows the prediction obtained with the exact Pauli operator in symmetric [left side Figs. (a) and (c)] and asymmetric [right side Figs. (b) and (d)] matter at a density equal to $k_{F_1} = k_{F_2} = 1.4 \units{fm^{-1}}$ and $k_{F_1} = 1.1 \units{fm^{-1}}, k_{F_2} = 1.4 \units{fm^{-1}}$ respectively.}
\label{Fig:np_E50}
\end{figure}

\begin{figure}[H]
\begin{center}
\subfloat{
\includegraphics[width=7cm]{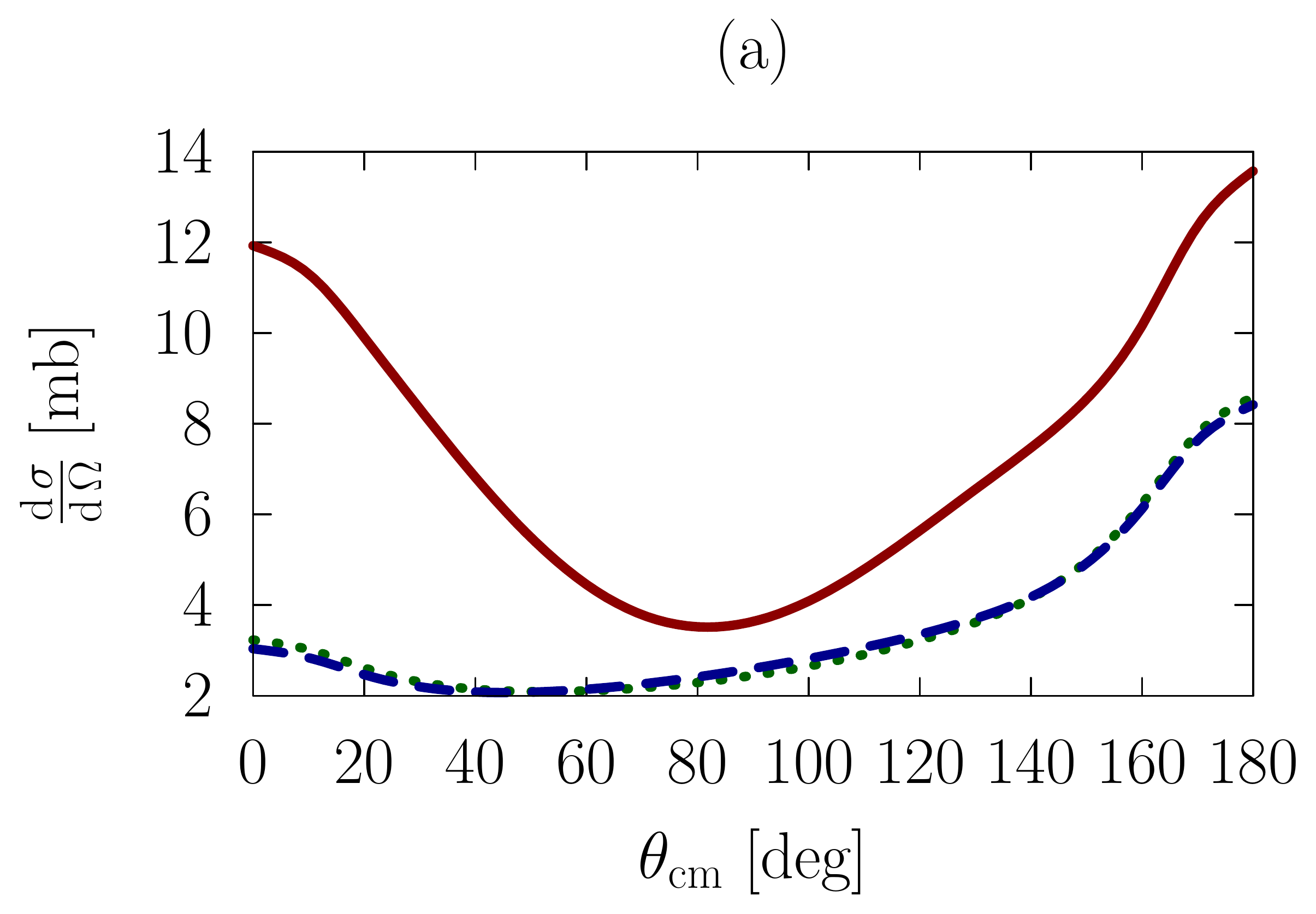}
}
\subfloat{
\includegraphics[width=7cm]{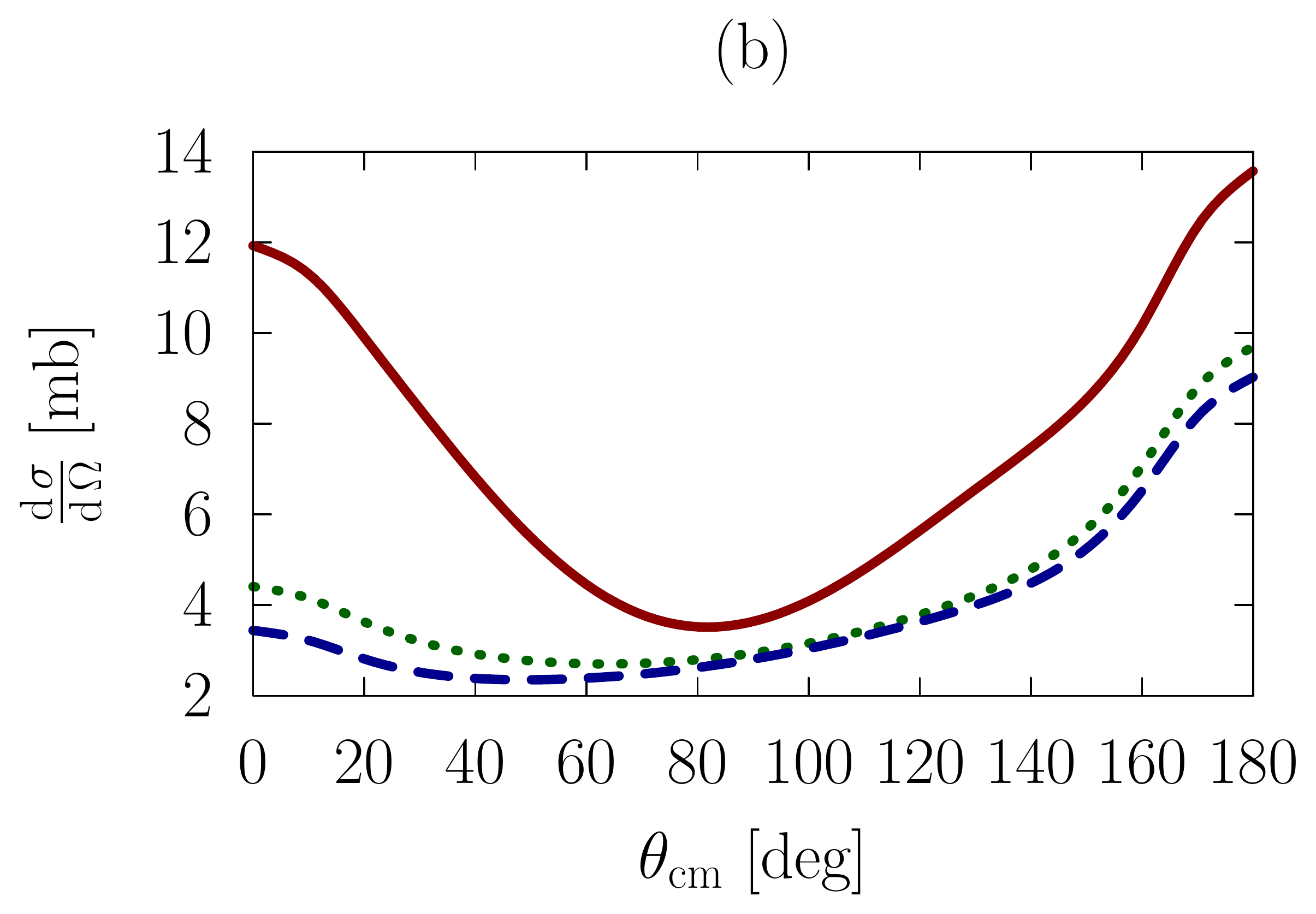}
}
\\
\subfloat{
\includegraphics[width=7cm]{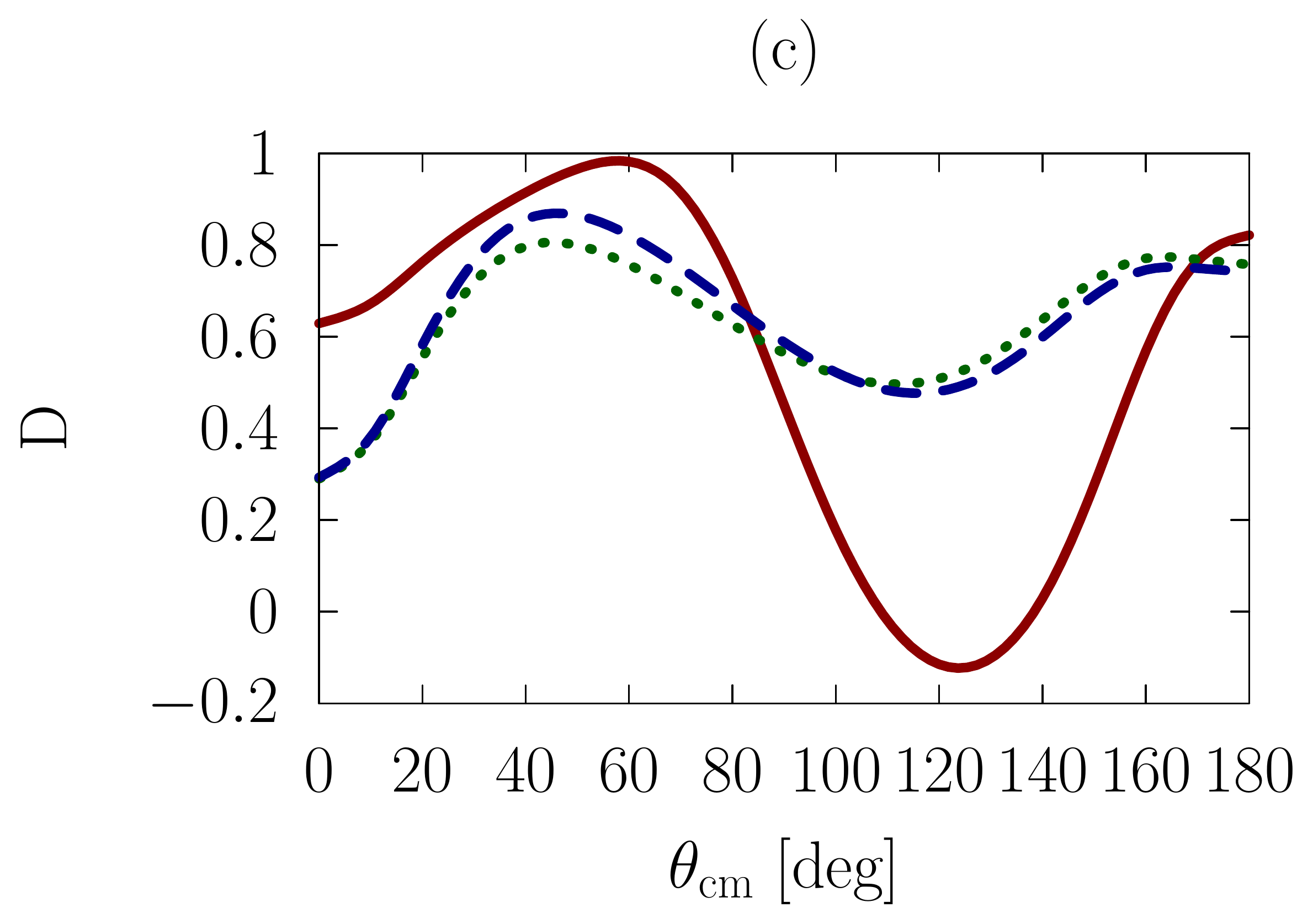}
}
\subfloat{
\includegraphics[width=7cm]{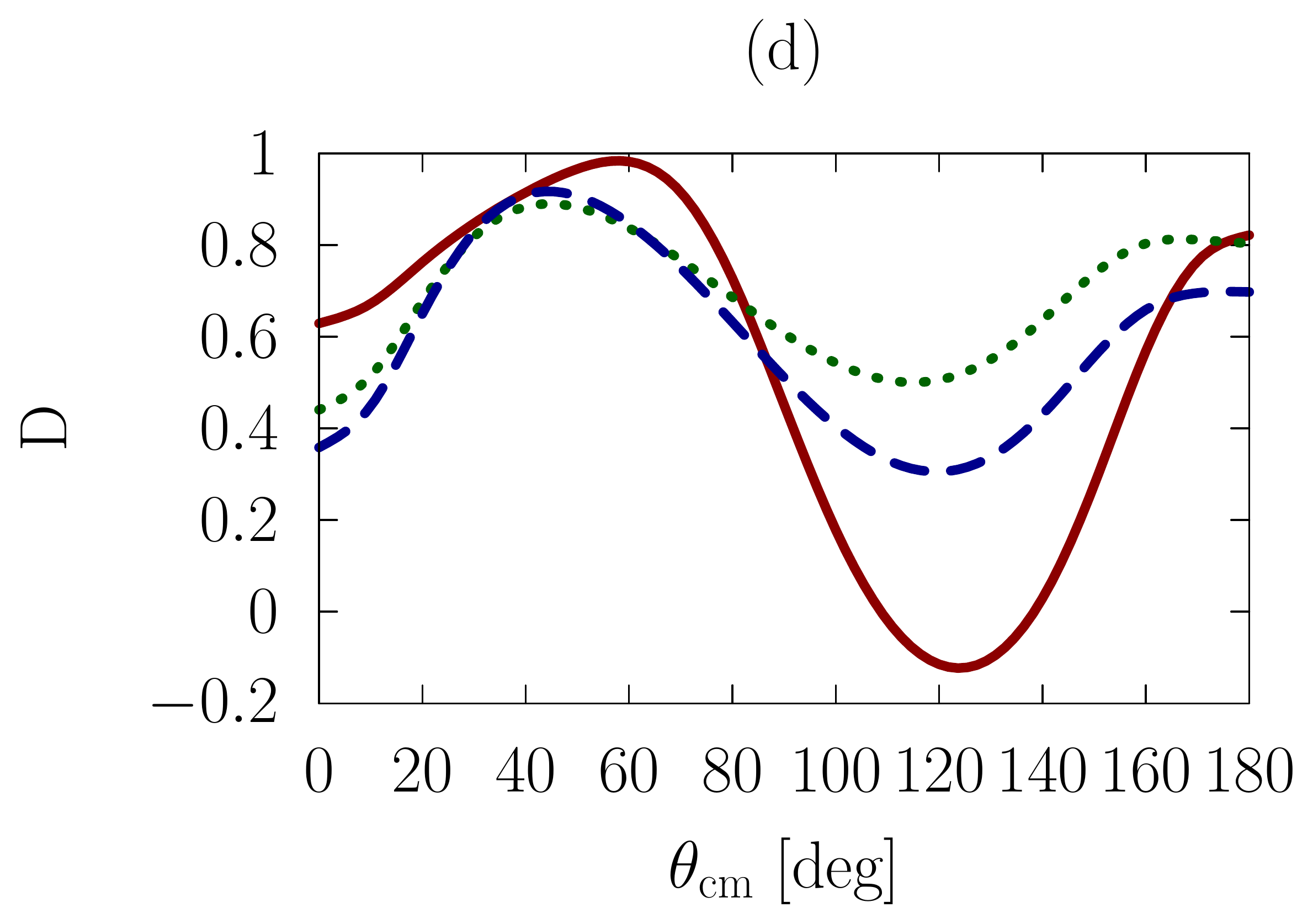}
}
\end{center}
\caption{(Color online) Same as Fig. \ref{Fig:np_E50} but at 100 MeV.}
\label{Fig:np_E100}
\end{figure}

\begin{figure}[H]
\begin{center}
\subfloat{
\includegraphics[width=7cm]{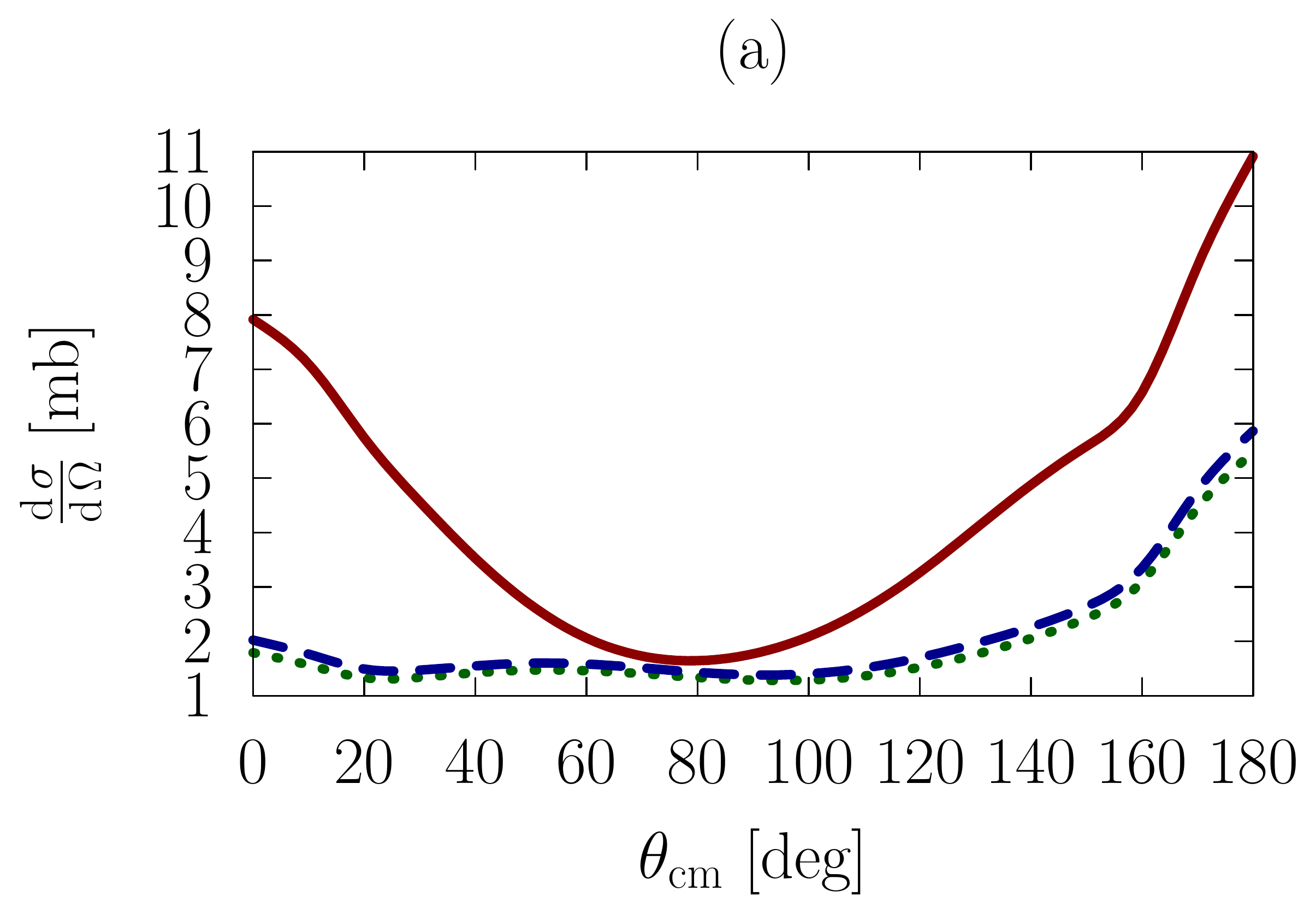}
}
\subfloat{
\includegraphics[width=7cm]{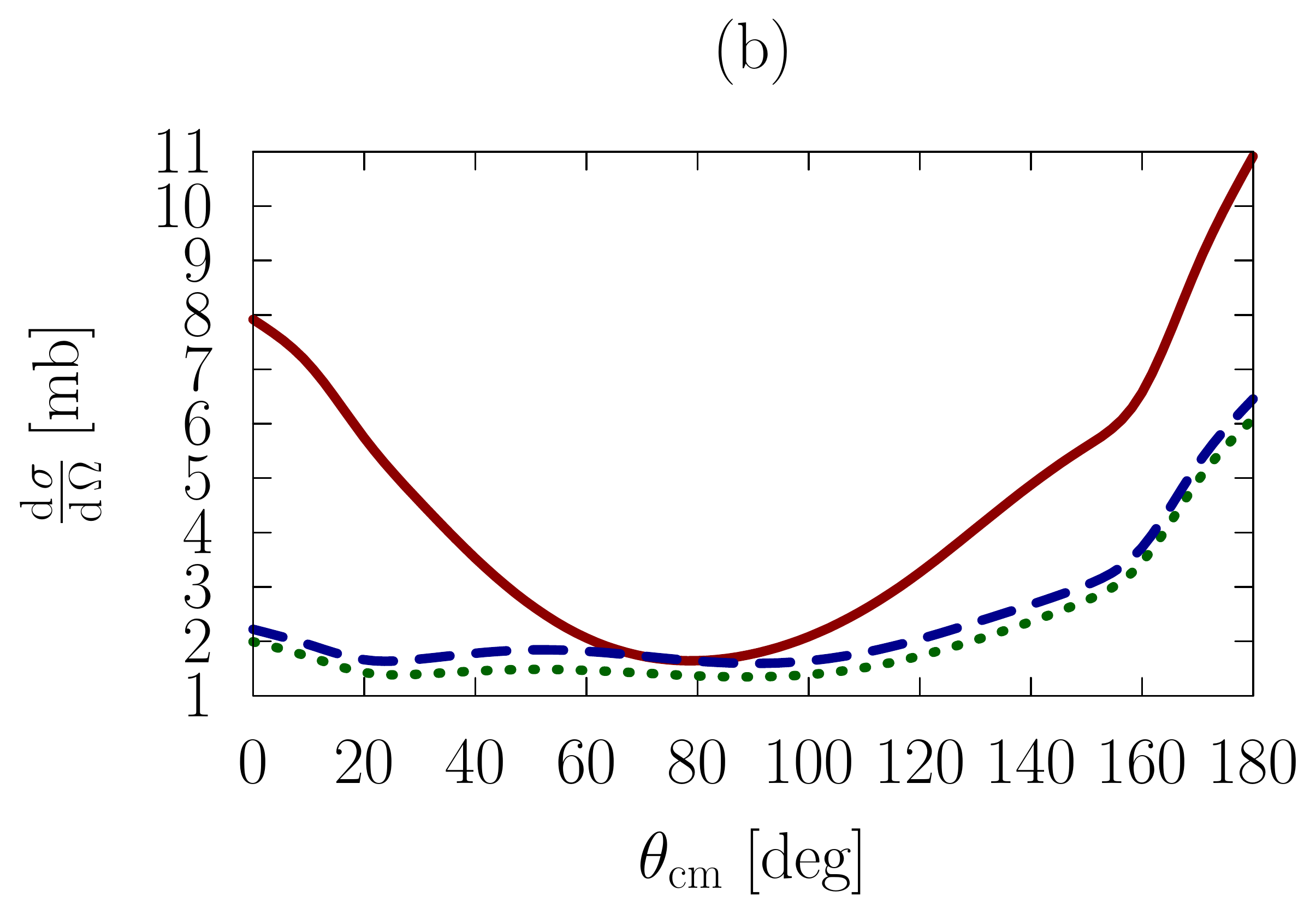}
}
\\
\subfloat{
\includegraphics[width=7cm]{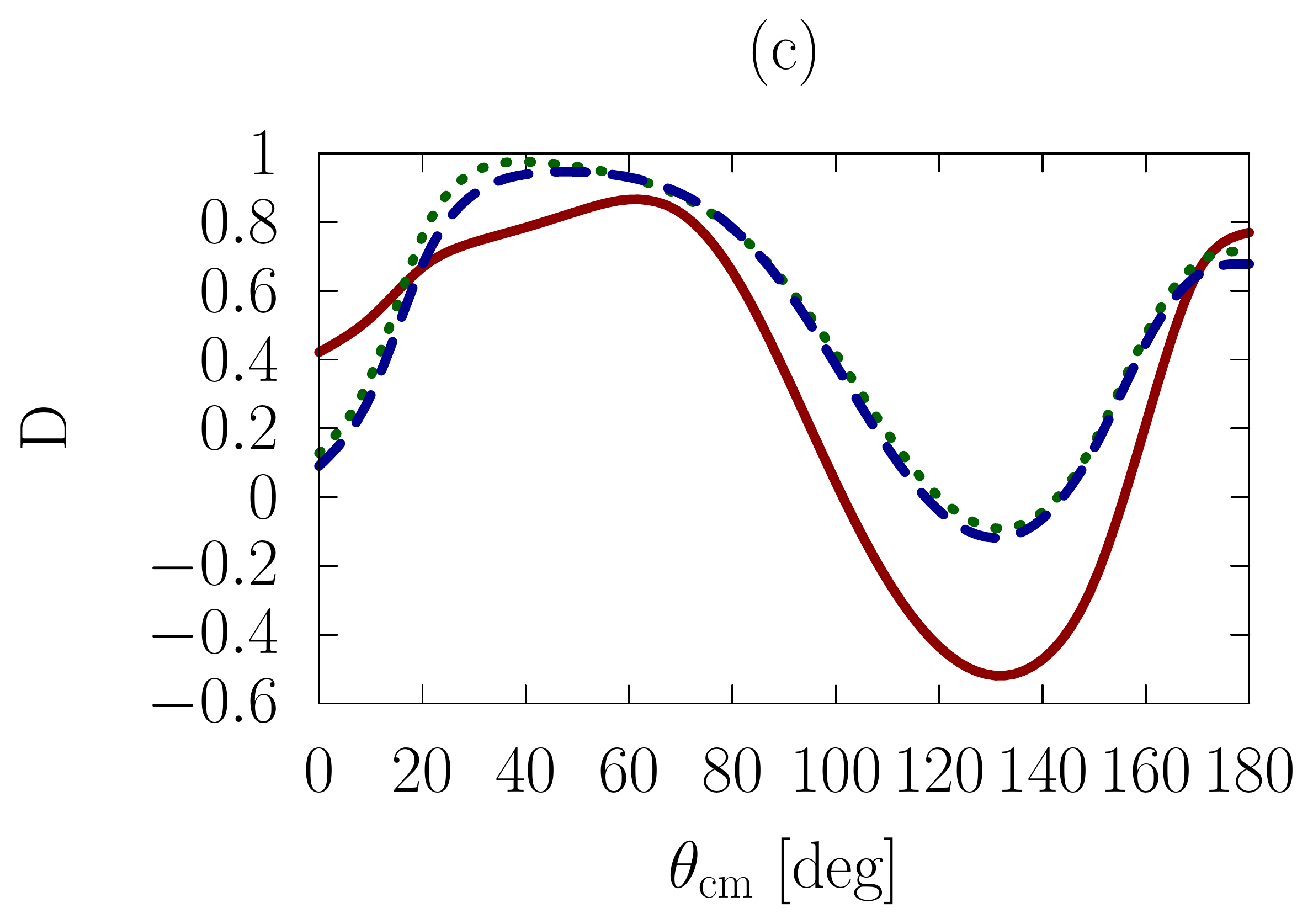}
}
\subfloat{
\includegraphics[width=7cm]{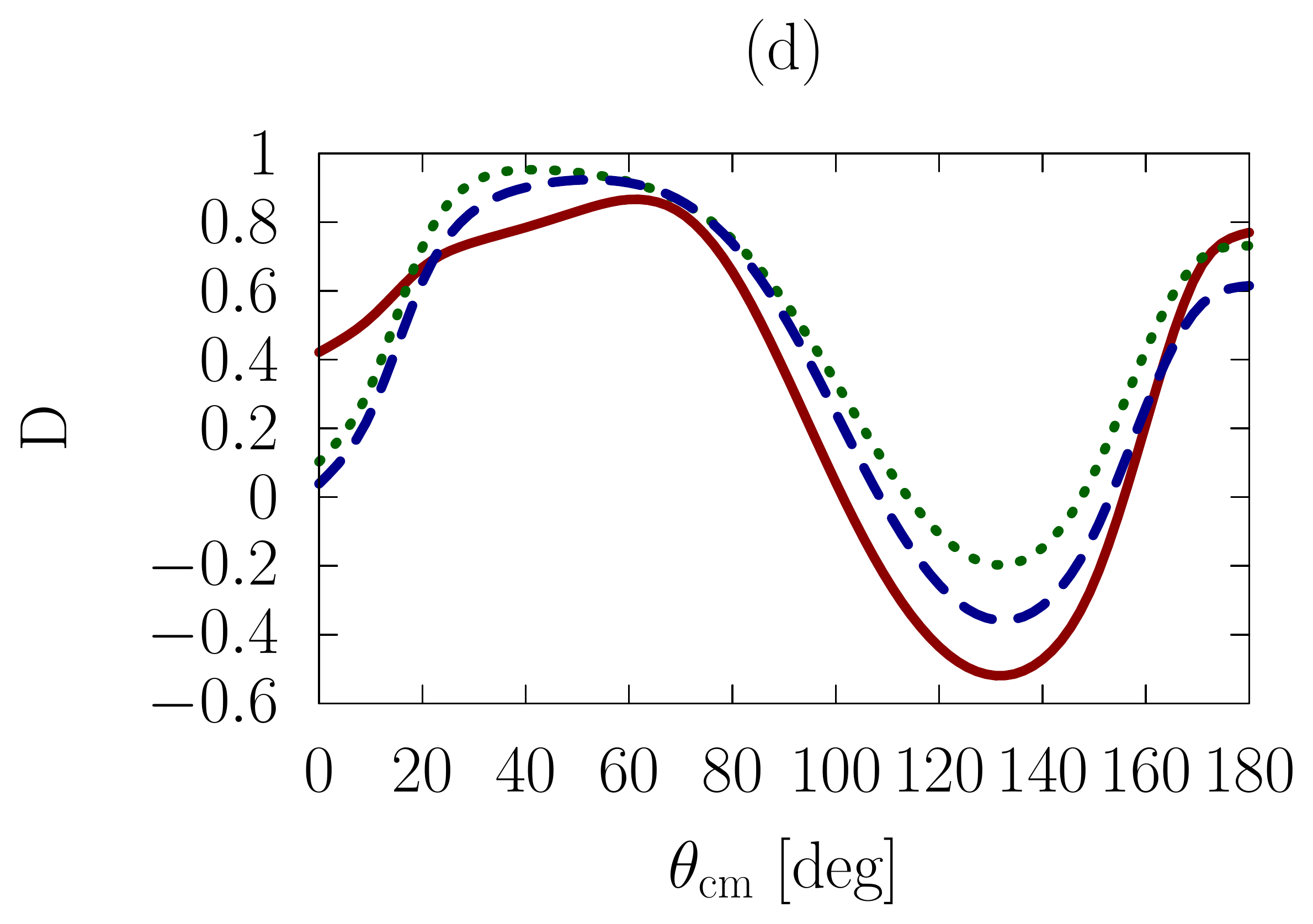}
}
\end{center}
\caption{(Color online) Same as Fig. \ref{Fig:np_E50} but at 200 MeV.}
\label{Fig:np_E200}
\end{figure}

In Figs.~\ref{Fig:np_E50}-\ref{Fig:np_E200}, we show $np$ observables at values of the on-shell c.m. momentum corresponding to
a free-space incident energy of 50, 100, and 200 MeV, respectively. In each figure, the frames
labeled as (a) and (b) display the elastic differential cross-section, whereas those labeled as
(c) and (d) show the depolarization parameter.

In all frames, the solid red curve shows the free-space predictions. For the frames on the left-hand side,
the dashed blue curve is obtained with the exact Pauli operator, assuming scattering in symmetric
nuclear matter with a Fermi momentum of $1.4 \units{fm^{-1}}$ [namely, we are solving the Bethe-Goldstone equation with the Pauli
operator as in Eq.~(\ref{Eq:Q})]; the dotted green curve is the corresponding result with the angle-averaged
calculation. For the frames on the right-hand side, we consider scattering in the presence of
two different Fermi momenta [see Eq.~(\ref{Eq:Q2})]. The dashed blue curve and the dotted green one are, again,
predictions with the three-dimensional formalism and the angle-averaged approach, respectively.

In all cases, medium effects on the energies are taken into account through the use of
nucleon effective masses, which we take from previous calculations \cite{Sam10}.
Specifically, the nucleon effective mass in nuclear matter with density corresponding to
$k_F=1.4 \units{fm^{-1}}$ is taken to be 612.8 MeV, whereas for
$k_F=1.1 \units{fm^{-1}}$ the value is found to be 718.3 MeV.

First, we observe that the density dependence is very large.
The differential cross-section is strongly reduced and flattened by medium effects.
Also, structures in the spin dependence are heavily suppressed.

Differences between the dashed curve and the dotted one are noticeable, but much
smaller than those between the free-space predictions and either one of the medium-modified calculations.
Interestingly, those differences are larger for the case of the asymmetric Pauli operator,
particularly so in the case of polarized scattering.

We note that the free-space $np$ cross-section is rather anisotropic, and becomes more so,
as energy increases, due to interferences from more partial waves.
In the presence of
medium effects, the cross-section becomes much more
isotropic. Also, medium effects are smaller at the higher energies, as is physically reasonable.

\begin{figure}[H]
\begin{center}
\subfloat{
\includegraphics[width=7cm]{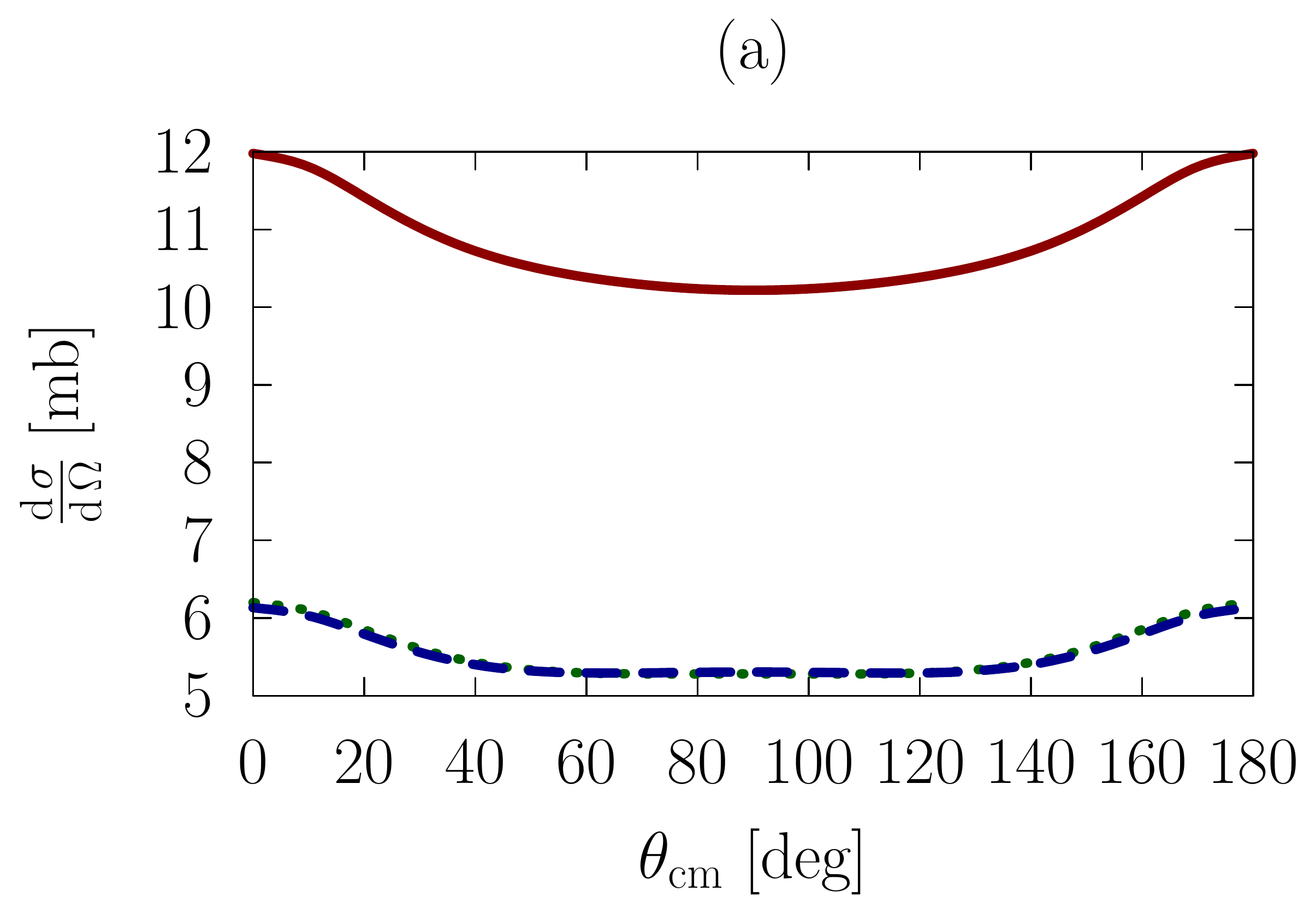}
}
\subfloat{
\includegraphics[width=7cm]{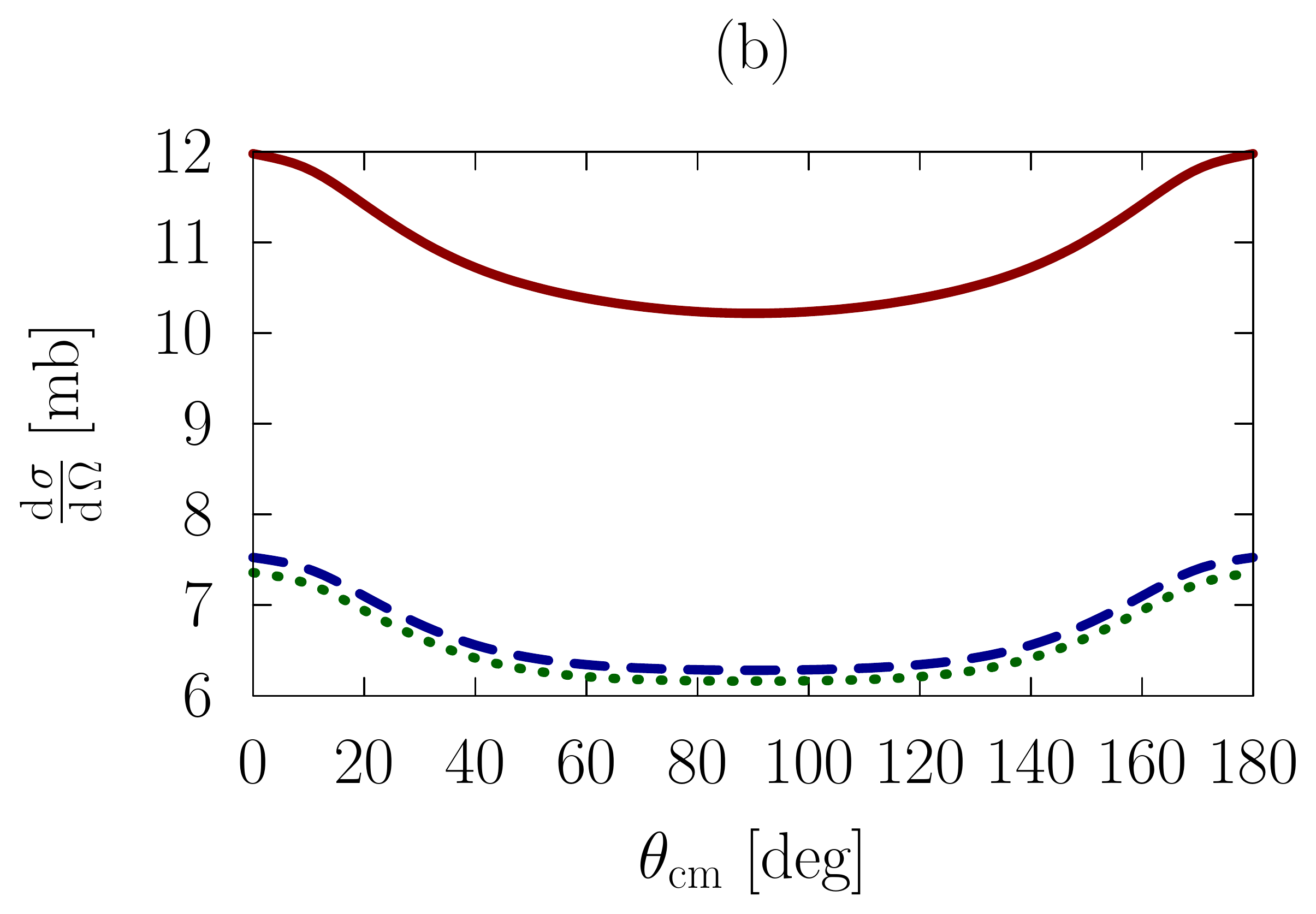}
}
\\
\subfloat{
\includegraphics[width=7cm]{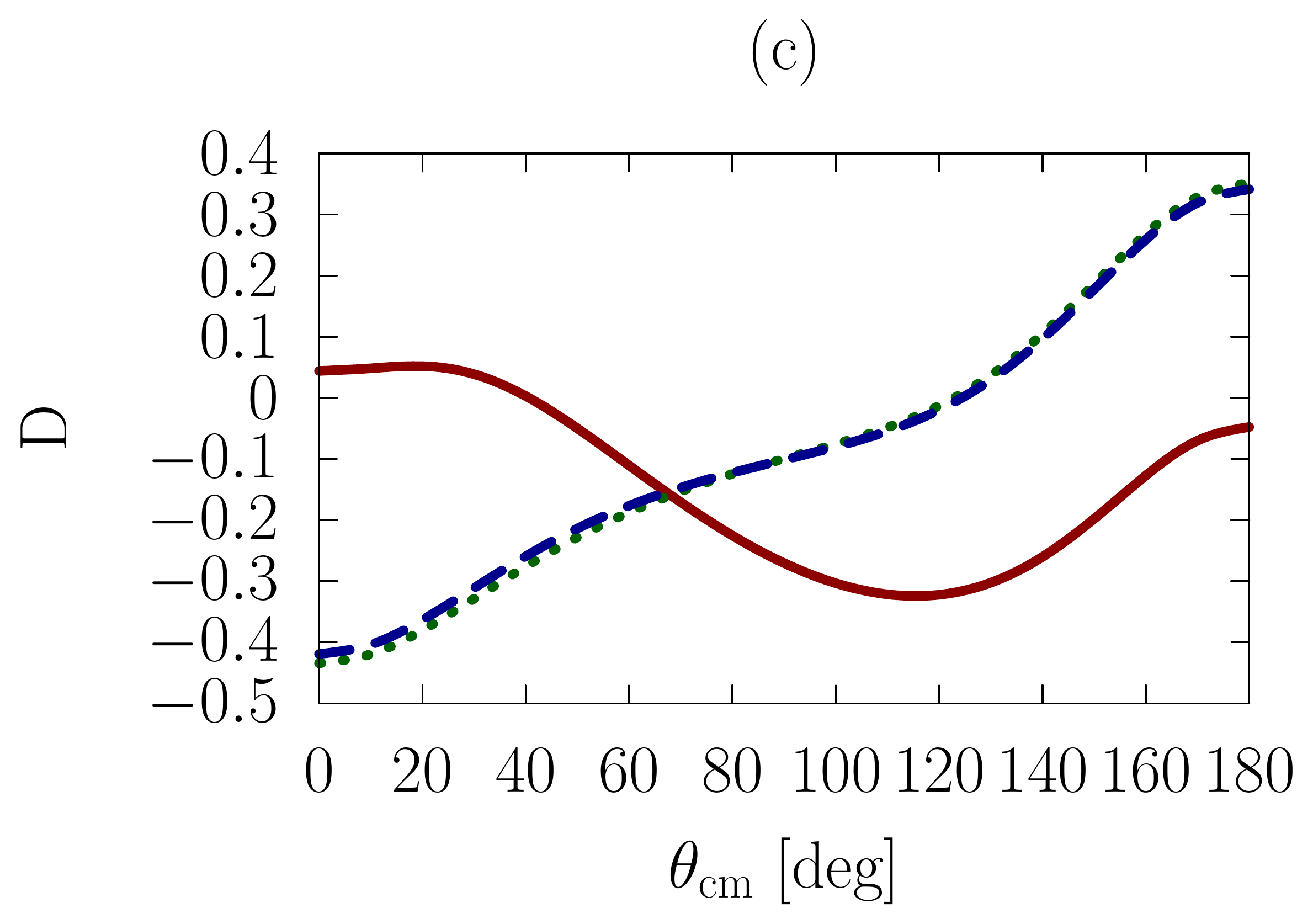}
}
\subfloat{
\includegraphics[width=7cm]{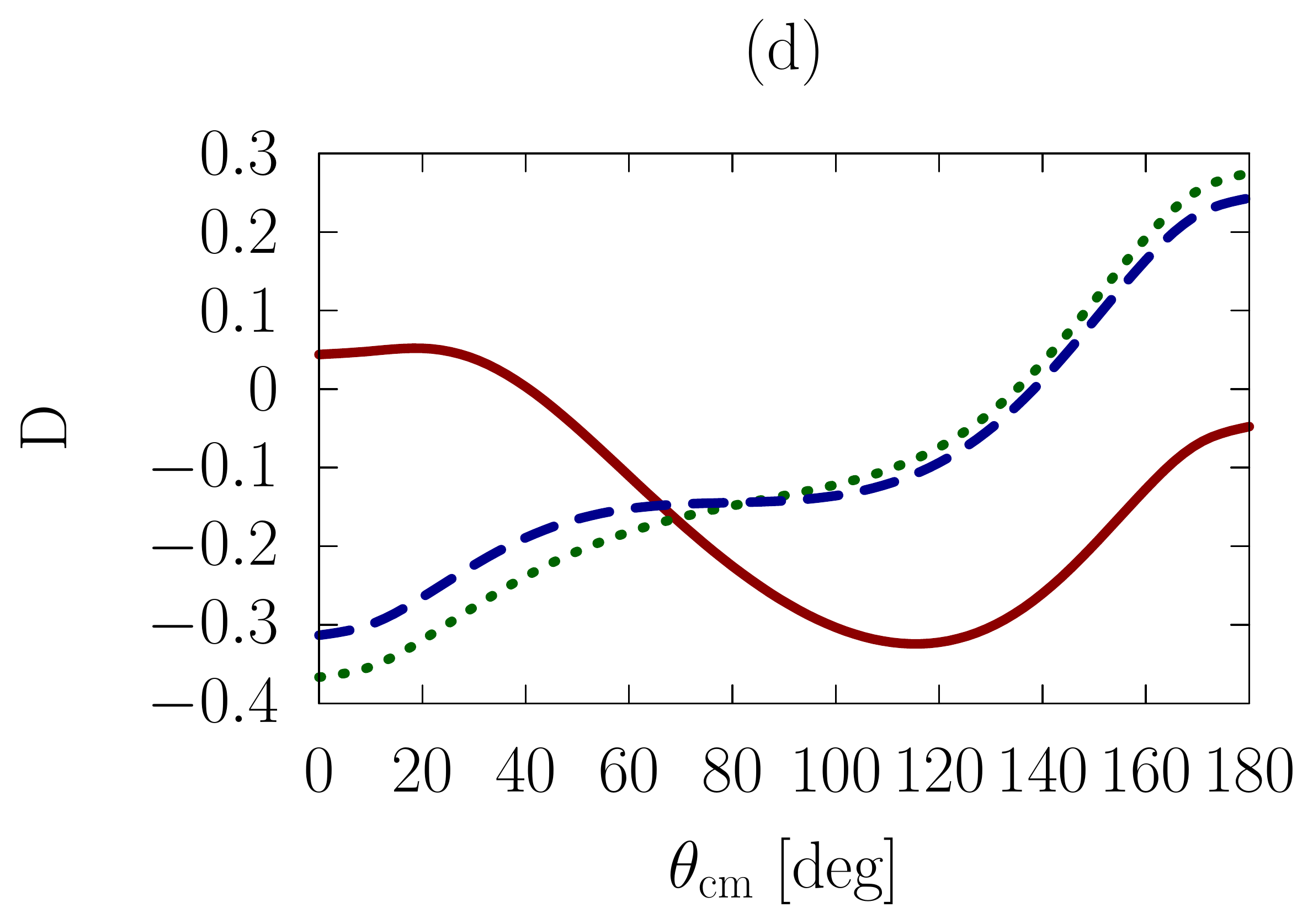}
}
\end{center}
\caption{(Color online) Same as Fig. \ref{Fig:np_E50} but for $pp$ scattering.}
\label{Fig:pp_E50}
\end{figure}

\begin{figure}[H]
\begin{center}
\subfloat{
\includegraphics[width=7cm]{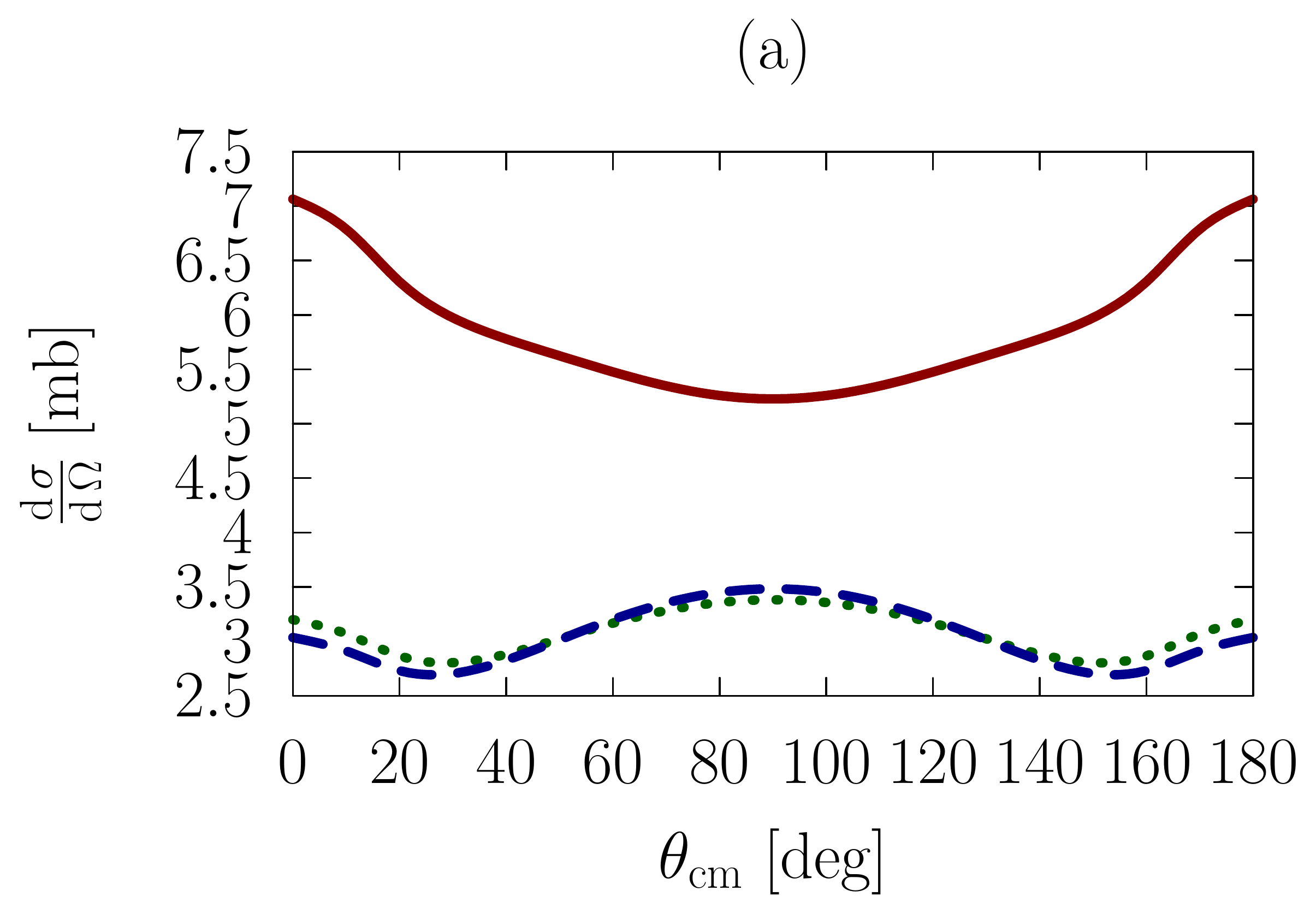}
}
\subfloat{
\includegraphics[width=7cm]{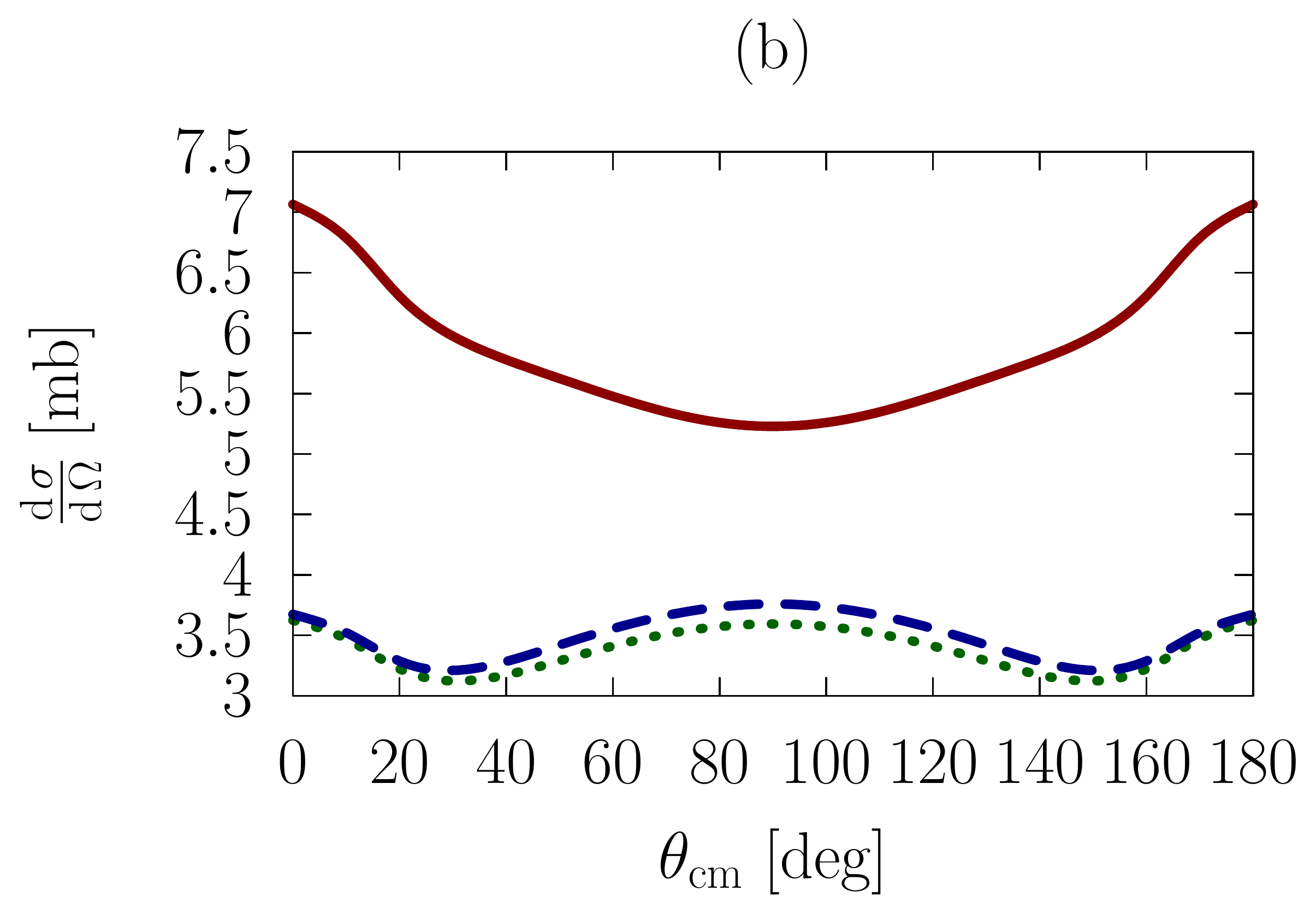}
}
\\
\subfloat{
\includegraphics[width=7cm]{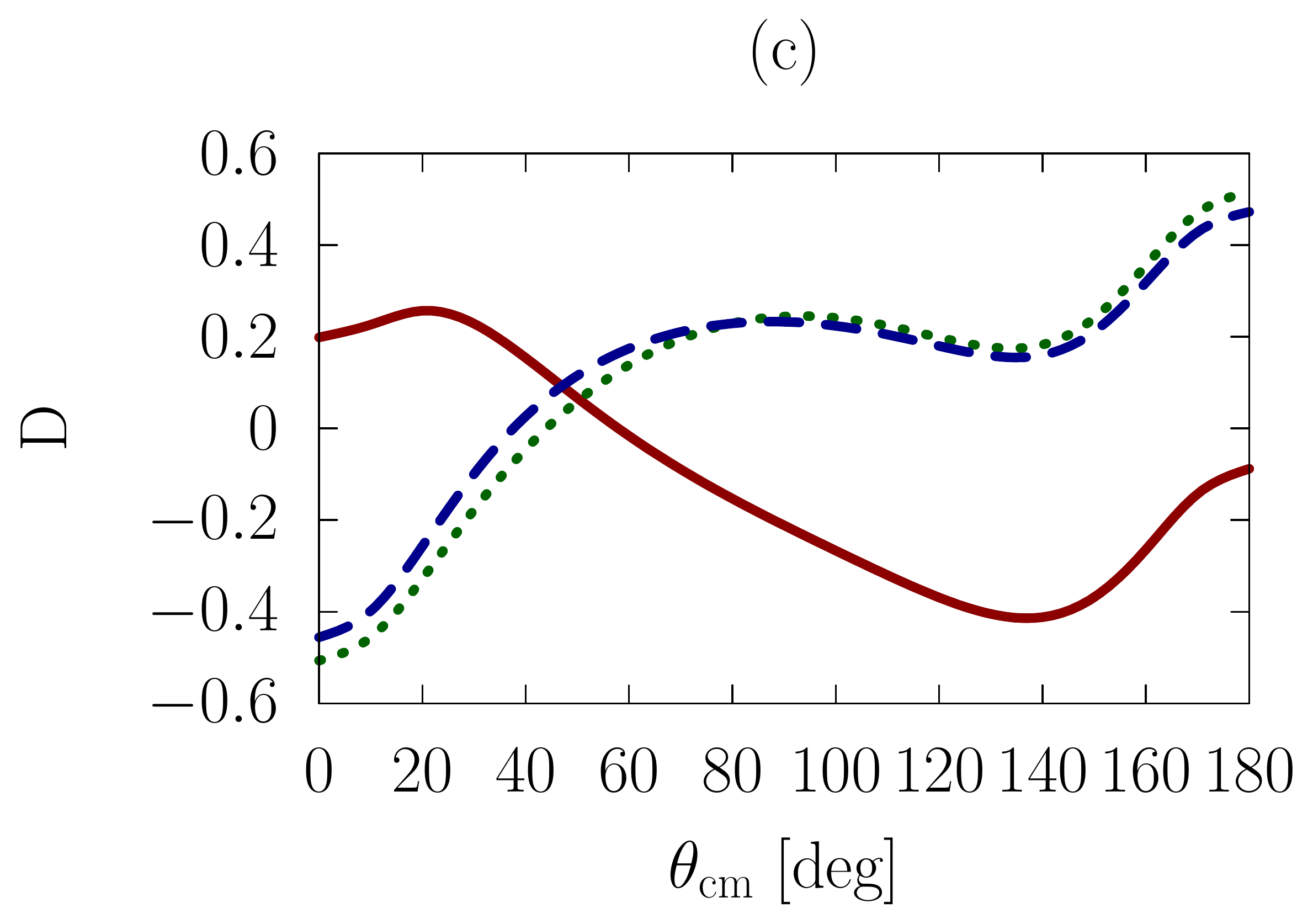}
}
\subfloat{
\includegraphics[width=7cm]{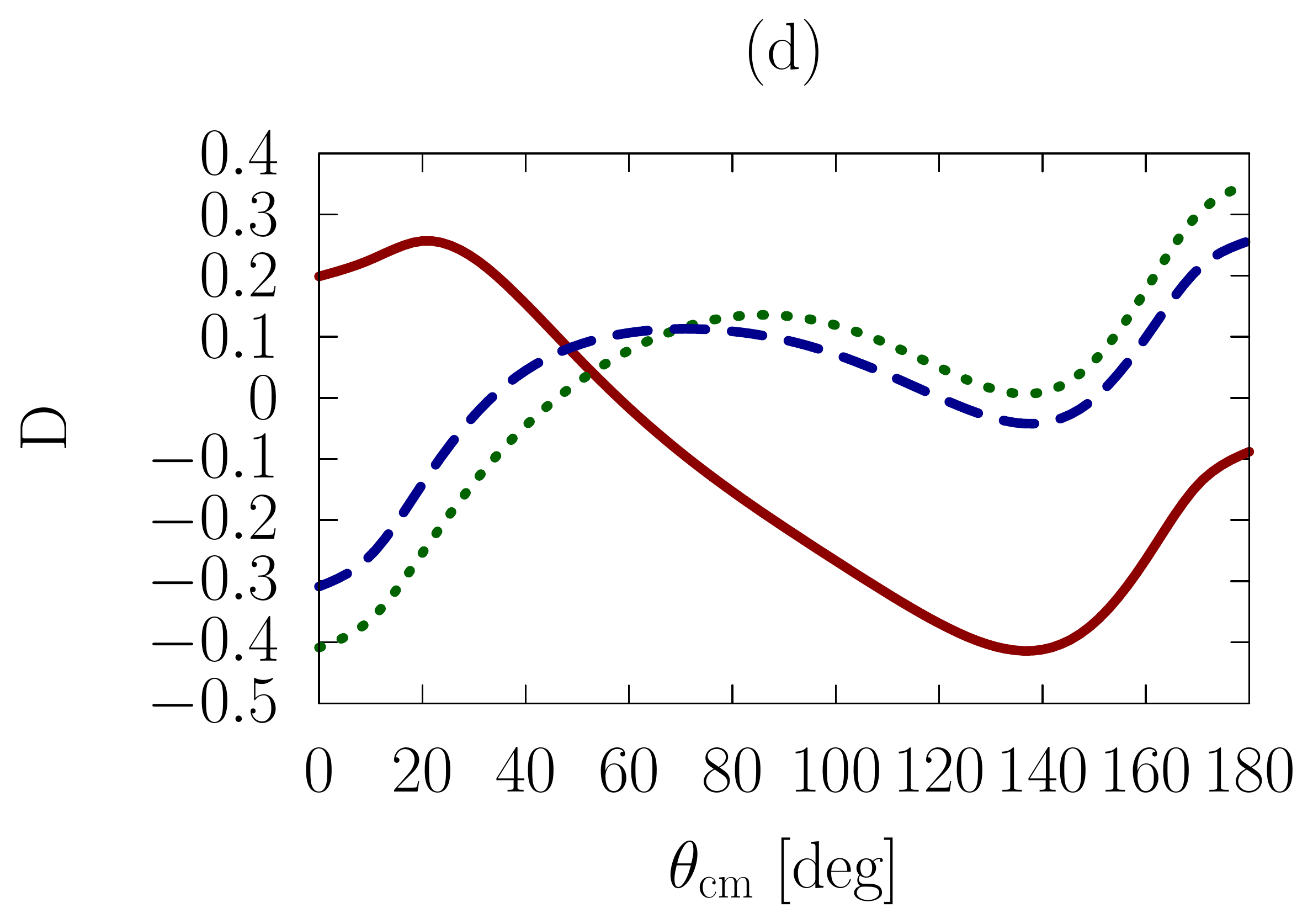}
}
\end{center}
\caption{(Color online) Same as Fig. \ref{Fig:pp_E50} but at 100 MeV.}
\label{Fig:pp_E100}
\end{figure}

\begin{figure}[H]
\begin{center}
\subfloat{
\includegraphics[width=7cm]{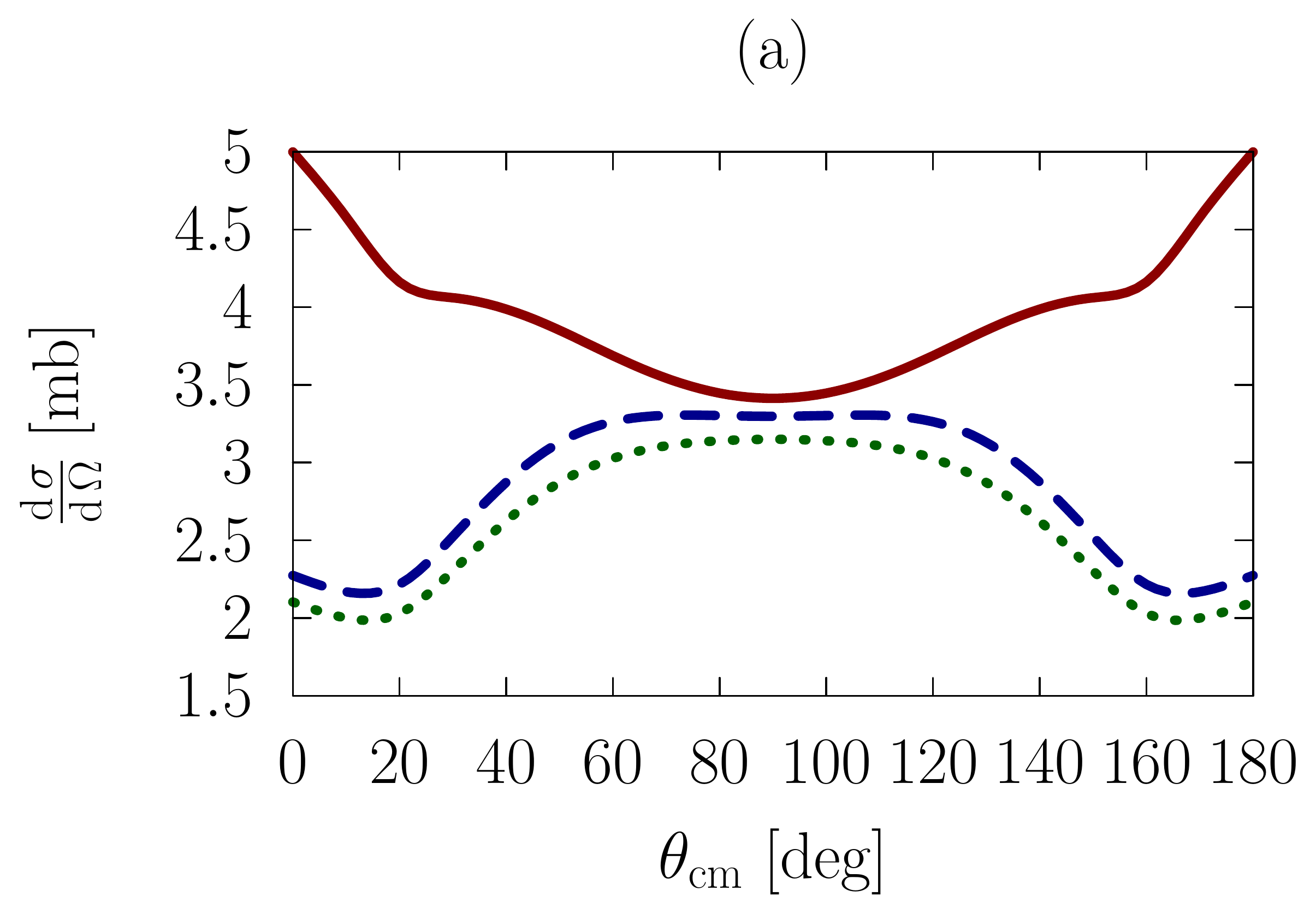}
}
\subfloat{
\includegraphics[width=7cm]{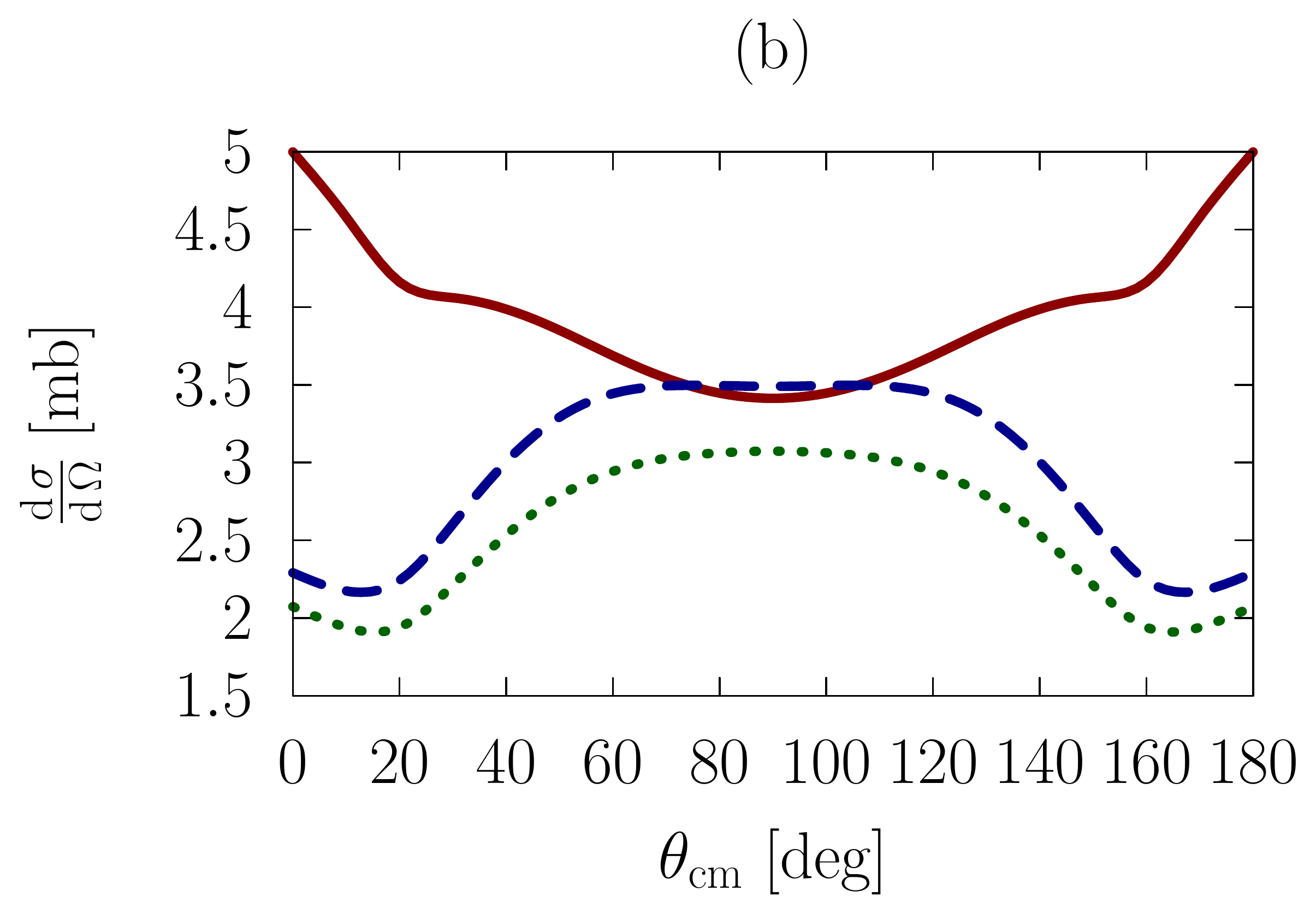}
}
\\
\subfloat{
\includegraphics[width=7cm]{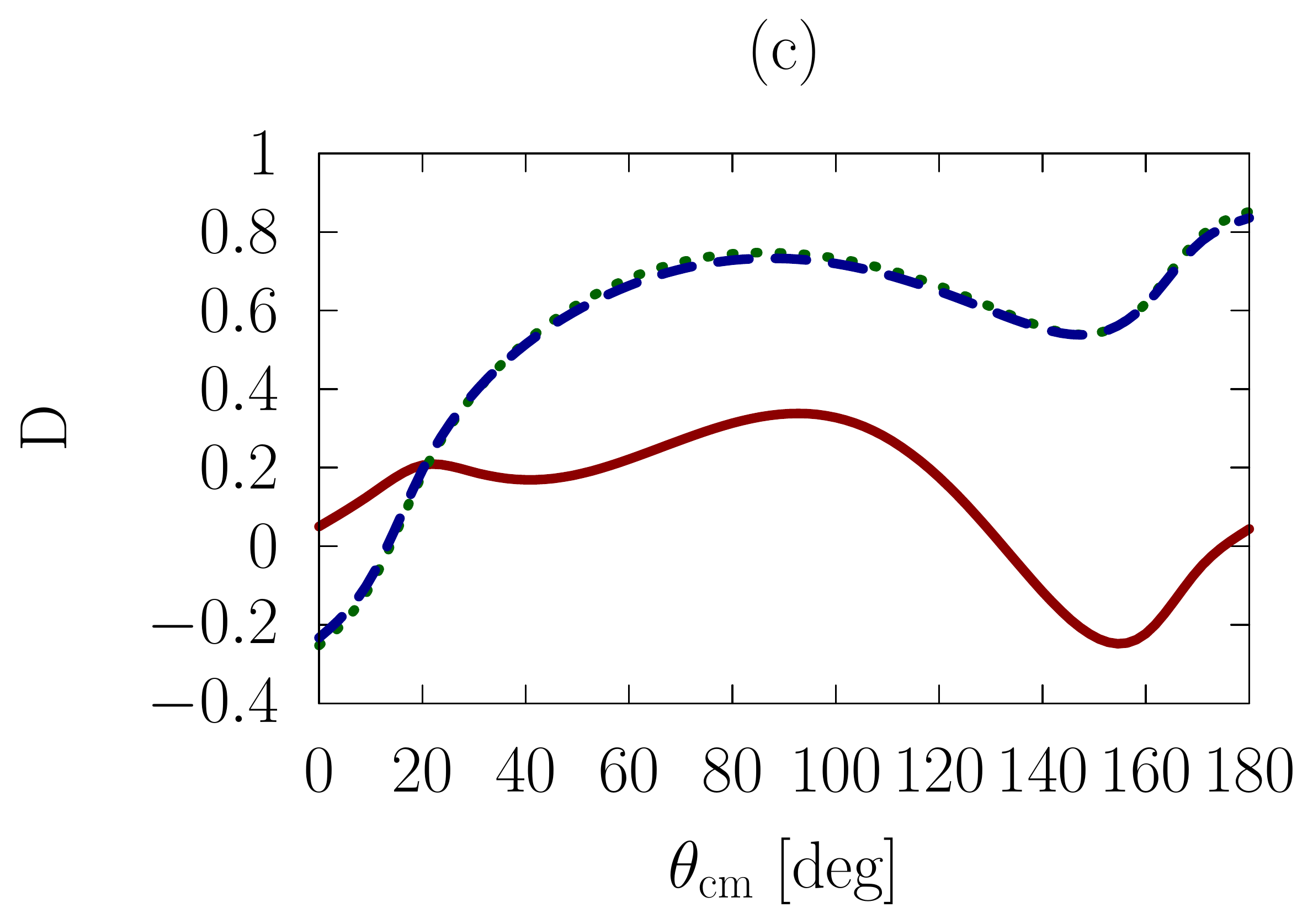}
}
\subfloat{
\includegraphics[width=7cm]{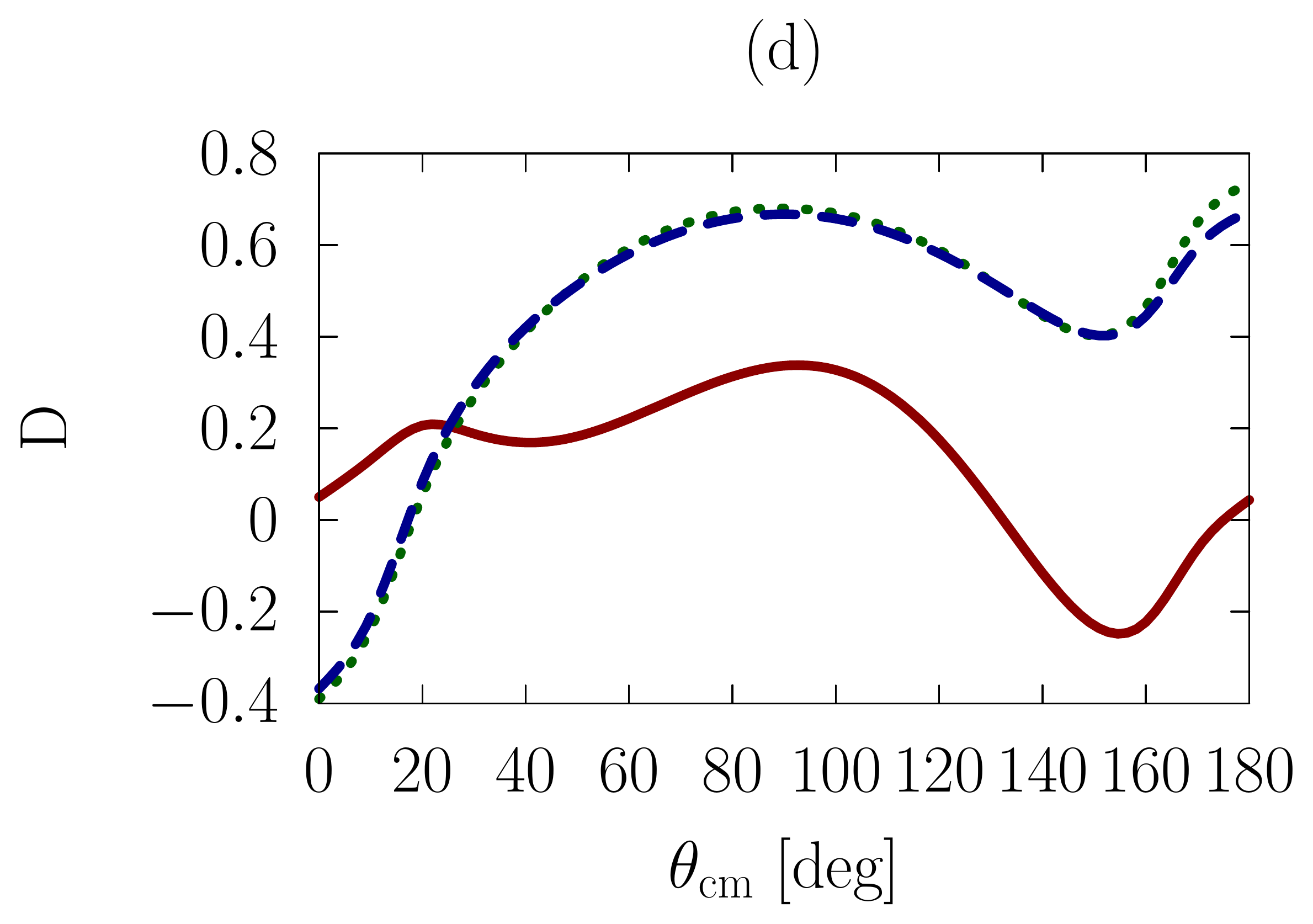}
}
\end{center}
\caption{(Color online) Same as Fig. \ref{Fig:pp_E50} but at 200 MeV.}
\label{Fig:pp_E200}
\end{figure}

In Figs.~\ref{Fig:pp_E50}-\ref{Fig:pp_E200}, we provide a similar presentation as the one in Figs.~\ref{Fig:np_E50}-\ref{Fig:np_E200}, but for
$pp$ scattering.
As far as general features are concerned, similar considerations apply. Namely,
there  is strong density dependence, and moderate sensitivity to the use
of the angle-dependent Pauli operator. Again, such sensitivity is more pronounced for the cases
on the left-hand side.

We note in passing that the free-space $pp$ differential cross-section is less anisotropic than
the $np$ one, due to the smaller number of partial waves that contribute to it ($I=1$ states only), and
is symmetric with respect to the $\theta \rightarrow \pi - \theta$ transformation.
In the medium, it is strongly reduced, and, at the higher
energies, shows a dramatic change in curvature. Recalling that different partial waves (beyond the $S$-waves) 
contribute non-isotropically to the cross section,
  we attribute this feature to how the relative role of different partial waves is impacted  by the medium .

With regard to sensitivity to the removal of the spherical approximation, we conclude that the latter is 
 slightly more pronounced in the $np$ case, particularly in the spin-dependent observable.
This suggests enhanced sensitivity in the $I=0$ channel, which is absent in the $pp$ interaction.

Overall, we can conclude that small effects  from the use of the non-averaged
Pauli operator are to be expected in applications
involving in-medium NN cross-sections.
Highly asymmetric situations (that is, two very unequal Fermi momenta), could be an exception.
Furthermore, it is appropriate to point out that the in-medium observables we have shown are obtained 
from the on-shell G-matrix. At this time, we cannot exclude a larger sensitivity to the removal of the spherical 
approximation in many-body calculations which utilize the half- or fully-off-shell G-matrix.

\section{Summary and Conclusion}\label{Summary}
We presented an alternative way to solve the Thompson and the Bethe-Goldstone 
equations in three-dimensional space. The main differences with the solution techniques developed 
in Sect.~\ref{Formalism} concern the way the azimuthal degree of freedom is integrated out in the equations and 
how the resulting analytical structure of the equations is handled. 

Exact angle-dependent on-shell amplitudes were calculated and used to obtain in-medium NN ``observables". These were compared to partial wave decomposed (angle-independent) solutions which utilized the spherical approximation on the Pauli operator. Only moderate sensitivity was observed between the exact and angle-averaged calculation in this case, although scattering in asymmetric matter (that is, in the presence of two different Fermi momenta), showed enhanced sensitivity.

Finally, Pauli blocking is one of the most important mechanisms
governing the scattering of fermions in the many-body system.
Regardless the magnitude of the effects we set forth to explore, to the best of our knowledge
the solution of the Bethe-Goldstone equation we have presented in this paper is an original one, and has allowed us to
better quantify the impact
of the historically very popular spherical approximation.

\begin{acknowledgments}
Support from the U.S. Department of Energy under Grant No. DE-FG02-03ER41270 is acknowledged.
\end{acknowledgments}
\bibliography{bibfile}
\end{document}